\documentclass[review,12pt]{elsarticle}

\usepackage{amssymb}
\usepackage{amsthm}
\usepackage{amsmath}
\usepackage{multirow}
\usepackage{tabularx}
\usepackage{mathrsfs}
\usepackage{graphicx}
\usepackage{epstopdf}
\usepackage{float}
\usepackage{caption}
\usepackage{subcaption}
\usepackage{bm}
\usepackage{upgreek}
\usepackage{bbm}
\usepackage{mathrsfs}
\usepackage{cleveref}
\usepackage{soul}
\usepackage{accents}
\usepackage{graphicx}
\usepackage{xcolor}
\usepackage{courier} 
\usepackage{listings} 
\usepackage{tabu} 
\usepackage{longtable}
\usepackage{changepage} 
\usepackage[margin=2cm]{geometry}
\biboptions{sort&compress} 

\journal{Additive Manufacturing}

\makeatletter
\def\@author#1{\g@addto@macro\elsauthors{\normalsize%
    \def\baselinestretch{1}%
    \upshape\authorsep#1\unskip\textsuperscript{%
      \ifx\@fnmark\@empty\else\unskip\sep\@fnmark\let\sep=,\fi
      \ifx\@corref\@empty\else\unskip\sep\@corref\let\sep=,\fi
      }%
    \def\authorsep{\unskip,\space}%
    \global\let\@fnmark\@empty
    \global\let\@corref\@empty  
    \global\let\sep\@empty}%
    \@eadauthor={#1}
}
\makeatother

\begin{document}

\begin{frontmatter}



\title{Hydrogen embrittlement susceptibility of additively manufactured 316L stainless steel: influence of post-processing, printing direction, temperature and pre-straining}


\author[IC,UO]{G. \'{A}lvarez}

\author[UVa]{Z. Harris}

\author[F]{K. Wada}

\author[UO]{C. Rodr\'{\i}guez}

\author[IC,Oxf]{E. Mart\'{\i}nez-Pa\~neda\corref{cor1}}
\ead{emilio.martinez-paneda@eng.ox.ac.uk}

\address[IC]{Department of Civil and Environmental Engineering, Imperial College London, London SW7 2AZ, UK}

\address[UO]{Department of Construction and Manufacturing Engineering, University of Oviedo, Gij\'{o}n 33203, Spain}

\address[UVa]{Department of Mechanical Engineering and Materials Science, University of Pittsburgh, Pittsburgh, PA 15261, USA}

\address[F]{National Institute for Materials Science (NIMS), 1-2-1 Sengen, Tsukuba, Ibaraki 305-0047}

\address[Oxf]{Department of Engineering Science, University of Oxford, Oxford OX1 3PJ, UK}

\cortext[cor1]{Corresponding author.}

\begin{abstract}

The influence of post-build processing on the hydrogen embrittlement behavior of additively manufactured (AM) 316L stainless steel fabricated using laser powder bed fusion was assessed at both room temperature and -50$^\circ$ C via uniaxial tensile experiments. In the absence of hydrogen at ambient temperature, all four evaluated AM conditions (as-built (AB), annealed (ANN), hot isostatic pressed (HIP), and HIP plus cold worked (CW) to 30\%) exhibit notably reduced ductility relative to conventionally manufactured (CM) 316L stainless steel. The AM material exhibits sensitivity to the build direction, both in the presence and absence of hydrogen, with a notable increase in yield strength in the X direction and enhanced ductility in the Z direction. Conversely, testing of non-charged specimens at -50$^\circ$ C revealed similar ductility between the CM, AB, ANN, and HIP conditions. Upon hydrogen charging, the ductility of all four AM conditions was found to be similar to that of CM 316L at ambient temperature, with the HIP condition actually exceeding the CM material. Critically, testing of hydrogen-charged samples at -50$^\circ$ C revealed that the ductility of the HIP AM 316L condition was nearly double that observed in the CM 316L. This improved performance persisted even after cold working, as the CW AM 316L exhibited comparable ductility to CM 316L at -50º C after hydrogen charging, despite having a 2-fold higher yield strength. Feritscope measurements suggest this increased performance is related to the reduced propensity for AM 316L to form strain-induced martensite during deformation, even after significant post-processing treatments. These results demonstrate that AM 316L can be post-processed using typical procedures to exhibit similar to or even improved resistance to hydrogen embrittlement relative to CM 316L.\\

\medskip

\end{abstract}


\end{frontmatter}


\noindent\textbf{Graphical abstract}

\medskip
\begin{figure}[h!]
\includegraphics[width=17cm]{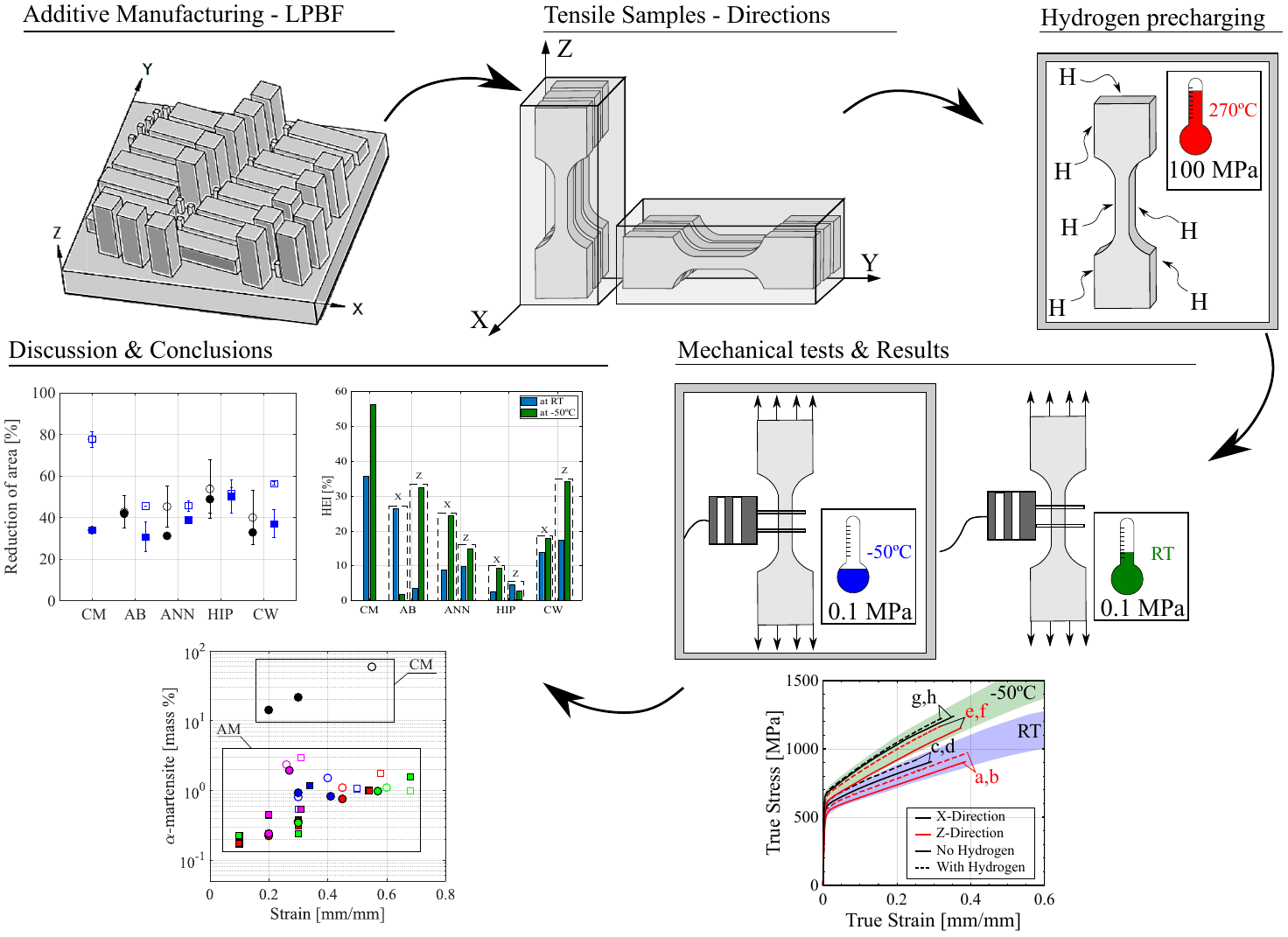}
\end{figure}

\section{Introduction}
\label{Introduction}
Metal additive manufacturing (AM) is a process in which components are fabricated from the `bottom up' using a layer-by-layer approach \cite{Frazier2014,Haghdadi2021}. This allows for geometrically complex components, such as those containing internal passages or novel geometric features, to be manufactured at near-net shape, thereby lowering the number of required machining steps, reducing material wastage, and decreasing production costs \cite{Savolainen2020,Singamneni2019,Simoes2022}. These benefits are most acute for components that are difficult to machine and/or highly specialized such that they are only made in limited quantities \cite{Singamneni2019,Haghdadi2021}. Additionally, since the physical footprint and utility requirements of AM systems are generally modest, they can be forward deployed (e.g., on-board naval craft or in difficult-to-reach worksites) to allow for on-demand production of replacement components with minimal lead time, enabling additional cost savings \cite{Sanchez2022,Kenney2013}. This opportunity for more time and cost-efficient production of specialized parts has motivated interest in leveraging AM to fabricate components used in challenging operating environments \cite{Singamneni2019,Sireesha2018,Sun2021,Vishnukumar2021,Blakey2021}. For example, AM has been proposed as a manufacturing strategy to produce components made from 316L stainless steel to support the hydrogen economy (e.g., nozzles \cite{Baek2017}, proton exchange membranes \cite{Kong2020}, and storage vessels \cite{Free2021}). It is well established that conventionally manufactured (CM) 316L typically exhibits good hydrogen compatibility \cite{Borchers2008,Dwivedi2018}. As such, there is significant interest in studying hydrogen effects on the mechanical behavior of AM 316L to assess whether a similar resistance to hydrogen embrittlement is observed.

In general, results have so far indicated that AM 316L exhibits similar mechanical behavior in air and after hydrogen charging, suggesting good hydrogen compatibility. For example, Khaleghifari et al. tested AM 316L fabricated via the laser powder bed fusion (LPBF) technique in the as-built condition after electrochemical charging, revealing a less than 5\% decrease in both the ultimate tensile stress and elongation to failure \cite{Khaleghifar2021}. A similar minor reduction in elongation to fracture was reported by Claeys et al. for LPBF AM 316L that was electrochemically charged with hydrogen \cite{CLAEYS2023}. This same resistance was also noted under gaseous environments, as Maksimkim et al. observed little change in the elongation to failure of AM 316L fabricated by LPBF when tested in high pressure (80 MPa) hydrogen gas \cite{Maksimkin2022}. However, it is important to note that the charging methods employed in these prior studies were insufficient to achieve tangible hydrogen ingress across the specimen thickness given the slow diffusivity of hydrogen in stainless steels ($\approx$1$\times$10$^{-16}$ m$^2$s$^{-1}$ \cite{Yamabe2017,Lin2020}). As such, rigorous interpretation of these data is obfuscated by the significant non-uniformity in hydrogen concentration (e.g., much of the specimen had a negligible hydrogen content).
  
The importance of testing methodology when assessing materials with high hydrogen resistance is underscored by the recent work of Bertsch et al. \cite{Bertsch2021}, who compared as-built and annealed AM 316L fabricated using both LPBF and directed energy deposition (DED). In their study, all specimens were hydrogen precharged using an elevated temperature gaseous hydrogen charging method \cite{Somerday2016}, where the samples are exposed to 120 MPa H$_2$ gas at 280$^{\circ}$ C for 400 hours to ensure that a uniform hydrogen concentration was achieved. Comparison of the four heat treatment/AM processing combinations revealed similar or improved mechanical behavior between the non-charged and hydrogen-charged specimens in the LPBF as-built, LPBF annealed, and DED annealed 316L conditions \cite{Bertsch2021}. However, the as-built DED 316L exhibited a 12\% reduction in the elongation to fracture after hydrogen charging \cite{Bertsch2021}. Interestingly, Hong et al. also observed a tangible debit in elongation to fracture after elevated temperature gaseous charging, though for as-built LPBF 316L ($\approx$25\% reduction) \cite{Hong2022}. This observation of inconsistently increased susceptibility strongly suggests an important role of alloy microstructure on the hydrogen compatibility of AM 316L. Such a dependence on alloy microstructure likely explains the variable performance of AM 316L relative to CM 316L when compared under similar environmental conditions. For example, Cruz et al. observed better resistance to stress corrosion cracking in as-built LPBF AM 316L relative to CM 316L during slow strain rate testing in 6 wt.\% FeCl{$_3$} solution \cite{Cruz2022}. Conversely, Miller et al. noted increasing susceptibility to hydrogen embrittlement in as-built LPBF AM 316L exposed to 0.75 M sulfuric acid solution for up to 90 days, while no effect was observed in the CM 316L \cite{Miller2018}. 
 
Mechanistically, the differences in hydrogen behavior between the CM and AM 316L (and likely between independent studies on AM 316L) has been attributed to the propensity (or lack thereof) for strain-induced martensite formation \cite{Kong2020,Metalnikov2022} or deformation twinning \cite{Hong2022} as well as the unique dislocation and grain structures that exist in the as-built AM 316L \cite{Bertsch2021,Lin2020,Hong2022,Khaleghifar2021,CLAEYS2023,Park2021}. However, it is notable that all studies comparing as-built and annealed 316L have revealed an increased hydrogen resistance after annealing \cite{Bertsch2021,CLAEYS2023}. It is well-established that annealing and other post-build processing (i.e., hot isostatic press (HIP) treatments) can significantly affect the mechanical and corrosion properties of AM 316L \cite{Shamsujjoha2018,Fonda2020,Afkhami2021,Vukkum2022,Chao2021, Grech2022,Hemmasian2018}. Critically, given that annealing will eliminate the unique as-built grain and dislocation cell structures, which have been proposed to promote deformation twinning \cite{Shamsujjoha2018,Pham2017,Liu2021,Hong2022}, this suggests that post-processing of AM 316L may improve both hydrogen compatibility and the consistency of alloy performance from build-to-build. However, dedicated studies examining the influence of post-build processing on the hydrogen performance of AM 316L are limited \cite{Bertsch2021,CLAEYS2023}, especially with regards to the influence of the commonly employed HIP treatment.

In addition to this identified knowledge gap regarding the influence of post-build processing on hydrogen compatibility in AM 316L, the influence of temperature on hydrogen embrittlement of AM 316L has also not been widely studied. Understanding the hydrogen susceptibility of candidate materials at cryogenic temperatures is important given the possibility that such low temperatures will be used in hydrogen storage systems \cite{Najjar2013}. Literature establishes that the hydrogen embrittlement susceptibility of austenitic stainless steels is actually highest at modestly cryogenic temperature (-100$^{\circ}$ C to -20$^{\circ}$ C) \cite{Borchers2008,Zhang2008,Zhang2013,Han1998,Michler2008}, with the extent of degradation dependent on the alloy composition \cite{Borchers2008,Kang2017,Mine2011,SanMarchi2008} and level of applied cold work \cite{Zhou2019,Michler2015}. The initial work of Hong et al. demonstrated that as-built LPBF 316L also exhibits increased hydrogen-induced degradation at 0$^{\circ}$ C and -40$^{\circ}$ C relative to ambient temperature \cite{Hong2022}. Critically, this increased susceptibility was related to the increased propensity for strain-induced martensite formation and deformation twinning, with the latter driven by the unique dislocation cell structure of the as-built material. As such, it is possible that annealing or HIP treatments may also improve the hydrogen performance of AM 316L under modest cryogenic temperature, but studies to date have yet to examine this possibility.

The preceding overview demonstrates that critical knowledge gaps exist regarding the influence of temperature and post-build treatment (such as HIP) on the hydrogen  compatibility of AM 316L. The objective of this study is to address these knowledge gaps by assessing the effect of different post-build processing methods on the hydrogen embrittlement behavior of AM 316L at ambient temperature and -50$^{\circ}$ C. Tensile experiments were performed both parallel and perpendicular to the build direction on non-charged and gaseously precharged AM 316L in four different processing conditions (as-built, annealed, HIP, and HIP plus cold worked). Results are compared to companion testing executed on CM 316L. The susceptibility to hydrogen embrittlement of the various AM conditions is then discussed in the context of observed microstructural features.

\section{Materials and Experimental methods}
\label{Sec:Methods}

\subsection {Materials}
\label{Sec:Materials}
Experiments were performed on both conventionally manufactured and additively manufactured 316L specimens. The CM 316L was obtained in the form of a 25-mm thick plate in the annealed condition. The AM 316L 'stock' material was printed as rectangular blocks with dimensions of 20 mm $\times$ 20 mm $\times$ 75 mm using 316L powder procured from Renishaw and a Renishaw AM400 laser powder bed fusion (LBPF) system at the US Army Research Laboratory. The long axis of the printed blocks was oriented either perpendicular (termed X direction) or parallel (Z direction) to the build direction, depending on the respective block. Pertinent build parameters include the use of 750 mm/s laser scanning speed, a stripe pattern with 5 mm-wide stripes and 60 $\upmu$m stripe overlap, layer thickness of 50 $\upmu$m, hatch distance of 110 $\upmu$m, point distance of 60 $\upmu$m, laser power of 195 W, and 80 $\upmu$s exposure time. These parameters were the best-practice build parameters for 316L according to Renishaw at the time of printing. No optimization of build parameters was performed for this study as the goal was to capture material behavior using supplier-suggested build parameters as would likely be implemented in service.  An Ar cover gas was used during the build and the printed blocks were arranged across the build plate in the same manner as Shoemaker et al. \cite{Shoemaker2022}. 

Once the build was complete, the blocks were cut from the build plate via electrical discharge machining (EDM). One subset of blocks were left in the as-printed state (referred to herein as the `as-built'; AB) and a second subset were vacuum-annealed at 1100$^{\circ}$ C for 1 hr, followed by cooling at 100$^{\circ}$ K/min (referred to as the `annealed' condition; ANN). The last subset underwent the following HIP treatment: a period of increasing temperature (10$^{\circ}$ K/min) and pressure (0.76 MPa/min) from ambient pressure and 100$^{\circ}$ C to 103.5 MPa and 1100$^{\circ}$ C, which was maintained for 200 minutes, followed by a decreasing temperature and pressure at the same rates of 10$^{\circ}$ K/min and 0.76 MPa/min. After the post-processing treatments were completed, tensile specimens with a 20-mm long gage length and 5-mm square gage cross-section were excised from the blocks using EDM. This sample geometry was based on ASTM E8 \cite{ASTME8}, but adjusted to optimize the AM build cost. Once the samples were machined, half of the HIP specimens were pre-strained to 30\% using a servohydraulic load frame at an initial strain rate of 3.2$\times$10$^-5$ $s^{-1}$; these samples are referred to as `CW' henceforth. A summary of the different post-processing conditions is provided in Table \ref{tableTreatments}.

\begin{table}[h]
\caption{Summary of the materials and treatments}
\label{tableTreatments}
    \begin{center}
    \begin{tabular}{p{0.1\linewidth} p{0.85\linewidth}}
    \hline
    ID & Treatment \\
    \hline
    {CM} & As-received plate in annealed condition \\ 
    {AB} & As-built by LBPF \\
    {ANN} & As-built by LBPF and then annealed at 1100$^{\circ}$ C for 1 hour \\
    {HIP} & As-built by LBPF and then HIP treated at 1100$^{\circ}$ C/103 MPa/200 min \\
    {CW} & As-built by LBPF, HIP treated at 1100$^{\circ}$ C/103 MPa/200 min, then pre-strained in tension to 30\% elongation \\
    \hline
    \end{tabular}
    \end{center}
\end{table}

The chemical composition of the CM and AM materials (AB condition) were measured using Optical Emission Spectroscopy (OES), with seven specimens tested for each material. The average compositions for each material are  shown in Table \ref{Chemical composition}.
All measured elemental concentrations were within the required ranges for 316L with the exception of the Ni content of the CM material (required range is 10-14 wt. \%).

\begin{table}[h]
 \centering
  \caption{Chemical composition (wt.\%)}
  \label{Chemical composition}
\begin{tabular}[t]{ccccccccc}
\hline
\text{Steel} & \text{Fe} & \text{C} & \text{Si} & \text{Mn} & \text{Cr} & \text{Mo} & \text{Ni} & \text{N}\\
\hline
CM & 69.49 & 0.022 & 0.282 & 1.80 & 16.6 & 2.02 & 9.08 & 0.003\\
AM & 67.99 & 0.018 & 0.604 & 0.57 & 16.3 & 2.40 & 11.7 & 0.003\\
\hline
\end{tabular}
\end{table}

\subsection {Hydrogen charging}
\label{Sec:Hydrogen Charging}
Once the post-processing treatments were completed, hydrogen was introduced into half of the specimens via gaseous charging. Given the low diffusivity of hydrogen in stainless steels \cite{Mine2011,Yamabe2017}, the samples were exposed to 100 MPa H$_2$ at 270$^{\circ}$ C for 400 hours. As shown in Appendix A, exposure to a temperature of 270$^{\circ}$ C during 400 hours does not bring any noticeable change to the mechanical behaviour. Established diffusion calculations \cite{Crank1975} indicate that this time is sufficient to introduce a homogeneous hydrogen concentration profile across the tensile specimen thickness. Once the specimens were removed from the charging autoclave, they were maintained at cryogenic temperatures to prevent hydrogen egress using either immersion in LN$_2$ (77 K) or storage in dry ice (195 K) until the tensile testing was conducted. 

The hydrogen content introduced during gaseous charging for the AB, ANN, and HIP treatments (Table \ref{tableTreatments}) was assessed via thermal desorption spectroscopy (TDS) conducted on witness flat plate coupons of similar thickness as the tensile specimens. TDS experiments were completed by heating each sample to 800$^{\circ}$ C at a constant heating rate of 100$^{\circ}$ K/h while under ultra-high vacuum conditions; a description of a typical TDS system is provided elsewhere \cite{Zafra2022}. The hydrogen egress was then monitored as a function of time (and therefore temperature) using an attached mass spectrometer system. The measured hydrogen egress versus time data was then integrated to determine the total hydrogen content. Representative values for the AB, ANN and HIP condition are 88, 100 and 87 wppm, respectively. These values are within the range of reported hydrogen absorption values for CM 316L under similar charging conditions that achieved a homogeneous hydrogen concentration throughout the sample (95-105 wppm, see \cite{SANMARCHI2021,NYGREN2021,SEZGIN2018,YAMABE2013}).

\subsection {Mechanical testing}
Tensile experiments on  the non-charged and hydrogen-charged specimens were carried out at both ambient temperature and -50$^{\circ}$ C using a servohydraulic load frame, with the specimen elongation actively measured through final fracture using an attached extensometer. For experiments conducted at ambient temperature, the hydrogen-charged specimens were removed from cryogenic storage and held at ambient temperature for 1 hour prior to the start of straining to allow for hydrogen redistribution, as discussed by Lassila et al \cite{Lassila1986}. All testing at ambient temperature was executed at a fixed displacement rate of 8$\times$10$^{-4}$ mm/s, which corresponded to an initial strain rate of 3.2$\times10^{-5}$ s$^{-1}$. The experiments at -50$^{\circ}$ C were completed using a frame-mountable MTS 651 environmental chamber system that allowed the controllable introduction of low temperature nitrogen gas from an LN$_2$ dewar into the environmental chamber to maintain a set temperature.

As with the ambient temperature testing, the hydrogen-charged specimens slated for testing at -50$^{\circ}$ C were removed from cryogenic storage and held at ambient temperature for 1 hour. They were then loaded into the environmental chamber and straining was initiated once the temperature was stabilized at -50$^{\circ}$ C.

Given that 316L exhibits significant ductility, even when charged with appreciable hydrogen concentrations \cite{SanMarchi2008}, the use of elongation to failure for comparing the hydrogen embrittlement performance of different conditions is compromised by extensive post-uniform deformation. As such, the reduction of area ($RA$) at the fracture location was  calculated using profile images taken with a calibrated optical microscopy (Zeiss Stemi 508 equipped with Axiocam 208 color camera) for each sample, as this metric provides a direct measurement of the true plastic strain at fracture \cite{Harris2018}. The reduction in area for the hydrogen-charged (RA$_H$) and non-charged (RA$_A$) specimens at a given condition were then used to calculate a hydrogen embrittlement index (HEI), defined as follows:
\begin{equation}
\label{eqn:HEI}
\text{$HEI$} (\%)=\frac{RA_{A}-RA_{H}}{RA_{A}}\times100
\end{equation}
\noindent i.e., a higher $HEI$ corresponds to an increased susceptibility to hydrogen embrittlement. 

\subsection {Microstructural analysis}
\label{Sec:Microstructural analysis}
Characterization of the material grain structure was performed via electron backscatter diffraction (EBSD) using a Zeiss Gemini scanning electron microscope (SEM) equipped with an Oxford EBSD system and operated at 20 kV. Samples were prepared for EBSD by grinding with SiC papers and then polishing with diamond slurries, finishing with 0.05 $\upmu$m collodial silica. The tilt of specimen to carry out the analysis was 70$^{\circ}$ and the step size of the scan was 1.1 $\upmu$m. EBSD data were analyzed using the open-source MTEX package for Matlab \cite{Bachmann2010}. In addition to EBSD, prepared samples were also evaluated at several magnifications using conventional scanning electron microscopy to assess porosity size distributions; $\geq$ 100 individual pores and lack of fusion (LOF) were used for subsequent analysis. SEM imaging was also performed on the fracture surface of each tested condition to understand how hydrogen and temperature affected the fracture mechanism. This fractography analysis was largely performed at the center of the specimen and utilized both low and high magnifications.

To complement the porosity characterization, the densities of the different materials were analyzed using Archimedes' method via water immersion. A total of six different measurements were done for each material. The results of this analysis are provided in the table below:

\begin{table}[h]
 \centering
  \caption{Densities in the different materials (mean $\pm$ standard deviation )}
  \label{Densities}
\begin{tabular}[t]{ccccc}
\hline
\text{Material} & \text{ {Density (g/cm$^3$)}}\\
\hline
CM & 7.982 $\pm$ 0.016 \\ 
AB & 7.967 $\pm$ 0.031 \\
ANN & 7.922 $\pm$ 0.049 \\
HIP & 7.910 $\pm$ 0.081 \\
\hline
\end{tabular}
\end{table}

The strain-induced martensite volume fraction was estimated using a Feritscope FMP30 and the 1.7$\times$ correction factor for calculating $\alpha$'-martensite content recommended by Talonen and coworkers \cite{Talonen2004}. Feritscope measurements were performed on the gauge length of the fractured specimens in a region away from the necking zone. It is assumed this location exhibited the maximum uniform strain measured from the stress-strain response of the sample. To understand how the strain-induced martensite content evolved with strain, several interrupted tensile experiments were performed on each material condition at both room temperature and -50$^\circ$. Note that the Feritscope measurements were performed after testing was completed (\textit{i.e.} at ambient temperature). These experiments were conducted in an identical manner as the tests to failure, but arresting the loading once a specific strain was reached. Feritscope measurements were then conducted on the gage section of each specimen. Note that all reported values of the $\alpha$'-martensite volume fraction are the average of three measurements and that error bars are not shown as they were smaller than the size of the plotted symbols (standard deviation below 0.67 and 0.06 for CM and AM respectively). To verify the values obtained with this method, X-ray measurements have additionally been carried out on some specimens. These measurements have been done with a Malvern Panalytical Empyrean X-ray diffractometer operating with CuK$\alpha$ (1.5406 \AA) at 45 kV and 40 mA. The scanning range 2$\theta$ was between 40-100$^\circ$  with a scanning rate of 0.01313$^\circ$. The results were analysed using the HighScore Plus software and a Rietveld analysis to obtain fraction of each phase.

\section{Results}
\label{Sec:Results}

\subsection {Microstructure}
\label{Sec:Results Porosity}
\begin{figure}[ptb]
    \centering
    \includegraphics[width=\textwidth]{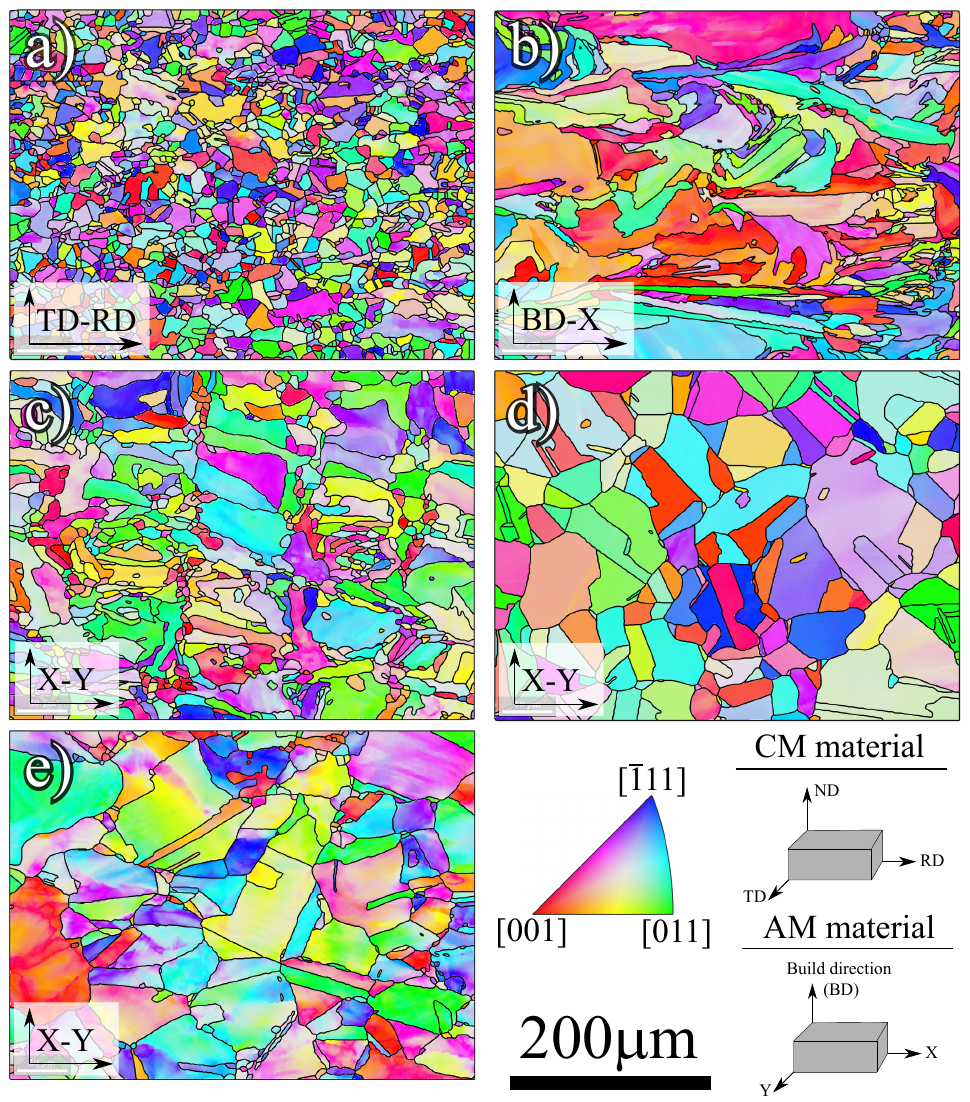}
    \caption{Overview inverse pole figure maps for each material condition: a) CM; b) AB; c) ANN; d) HIP; e) CW.}
    \label{fig:EBSD}
\end{figure}

Inverse pole figure (IPF) maps of the grain structure for each evaluated material in the non-strained condition are shown in Figure \ref{fig:EBSD}. These EBSD data (as well as collected X-ray diffraction results) confirm that all conditions exhibited a typical single phase austenitic microstructure, with the exception of the CM condition, which contained ferrite stringers oriented along the rolling direction of the plate as other authors have observed for this type of material \cite{Benrhouma2017,Ferrandini2006}. The CM 316L (Figure \ref{fig:EBSD}a) revealed a nominally equiaxed grain structure (grain size of $\approx25 \, \upmu$m), consistent with expectations for an annealed CM material, while the AB 316L (Figure \ref{fig:EBSD}b; with Y-Z being the plane of examination) exhibits a directional, irregular grain structure (grain size of $\approx15.5 \, \upmu$m) that is similar to the reported microstructure of AB 316L fabricated via LPBF \cite{Shamsujjoha2018}. Interestingly, the 1 hr annealing treatment was not sufficient to induce an equiaxed grain structure in the ANN 316L specimen, as shown in Figure \ref{fig:EBSD}c (grain size similar to AB material). However, an equiaxed structure was recovered after the prolonged HIP treatment (Figure \ref{fig:EBSD}d), which also resulted in a coarse grain size of $\approx75 \,\upmu$m. Lastly, as expected given the use of the same HIP treatment, the CW condition exhibits a coarse, equiaxed grain structure, though notable variations in orientation are observed within the grain interiors (Figure \ref{fig:EBSD}e). Such variations in the intragranular orientation are generally attributed to plastic deformation processes \cite{Bay1992,Brewer2009}, consistent with the prestrained nature of the CW specimens.

In addition to evaluating the different grain structures for the tested materials, it is important to recognize that AM materials can exhibit defects unlikely to be observed in CM alloys. For example, porosity has been widely reported in AM materials \cite{Yin2019} and can negatively impact mechanical performance. The nature and origins of the porosity can vary, with sources including lack-of-fusion (LOF) defects, gas entrapment, moisture content, and keyhole-driven porosity. Notably, the morphology and distribution of the pores can be modified by applied post-processing steps, therefore it is important to directly evaluate these features for each considered post-processing condition. Representative micrographs of porosity in each evaluated CM and AM material condition are shown in Figure \ref{fig:Porosity}. Additional images from other planes in the AM condition are provided in Appendix C for reference. The CM 316L contains a sparse population of void-like features (example indicated by the white arrow in Figure \ref{fig:Porosity}a), but these are relatively infrequent compared to the widespread defects observed in the AM materials. However, as expected from literature \cite{Yin2019}, the character of these defects notably changes with post-processing. For example, the AB and ANN 316L exhibit pore-like features, microcracks, and unmelted particles, with examples of the latter two features in the AB 316L indicated by the white arrows in Figure \ref{fig:Porosity}b. Large-scale, jagged pore-like features approaching tens of micrometers in size are also observed in the AM conditions that did not under go a HIP treatment; an example of these features from the ANN 316L condition is shown in Figure \ref{fig:Porosity}c (white arrows). However, after the HIP treatment is applied, cracks, pores and LOF are largely absent from the microstructure and the primary defect becomes widespread sub-micrometer size porosity (indicated by the white arrows in Figure \ref{fig:Porosity}d-e for the HIP and CW conditions). These small-scale porosity has been previously reported \cite{Shoemaker2022} and are likely due to trapped Ar from the build process.

\begin{figure}[p]
    \centering
    \includegraphics[width=\textwidth]{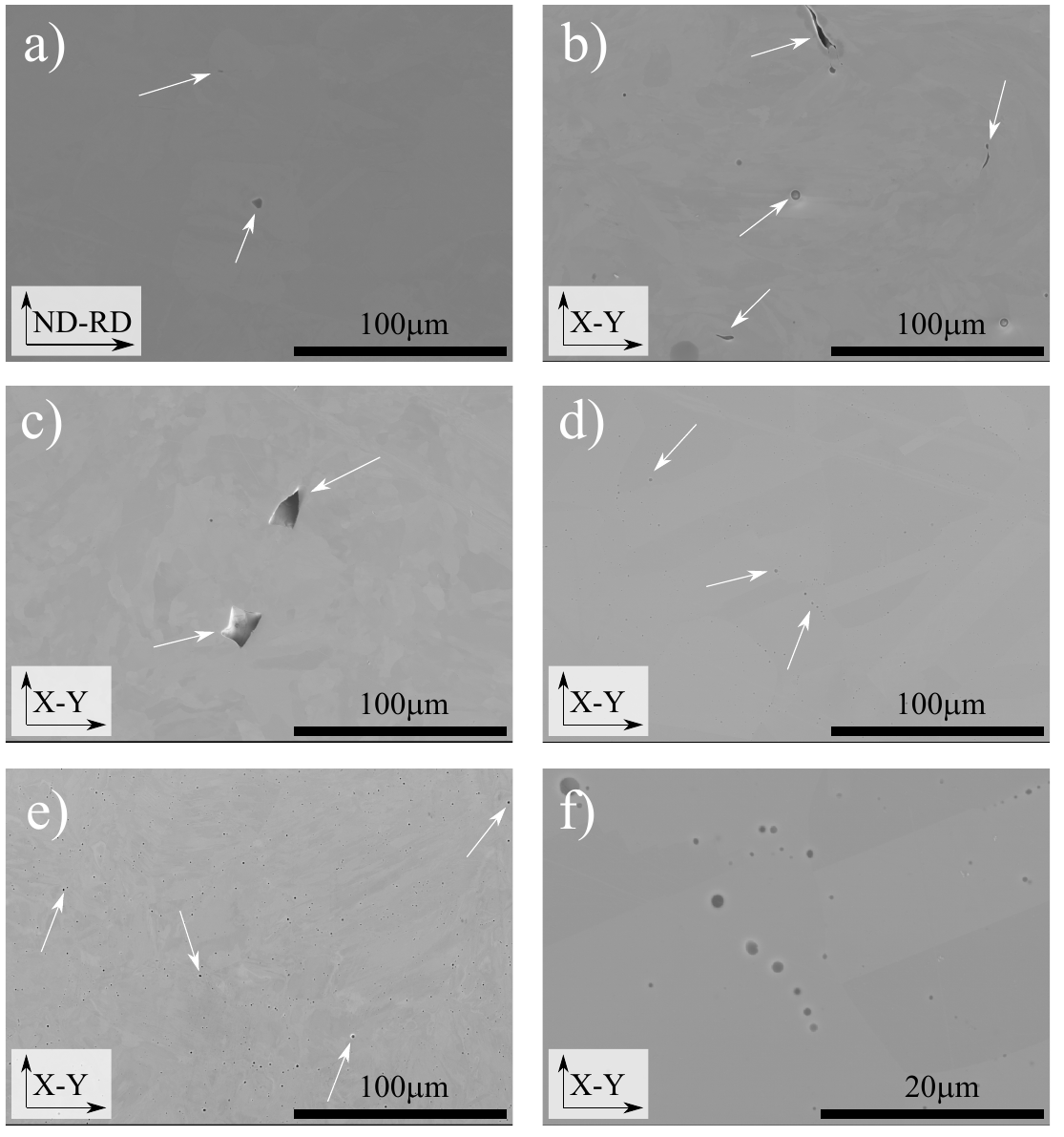}
    \caption{Porosity and defects in the different material conditions: a) CM; b) AB; c) ANN; d) HIP; e) CW; f) higher magnification images of the pores shown in the HIP and CW materials.}
    \label{fig:Porosity}
\end{figure}

\begin{figure}[ht]
    \centering
    \includegraphics[width=\textwidth]{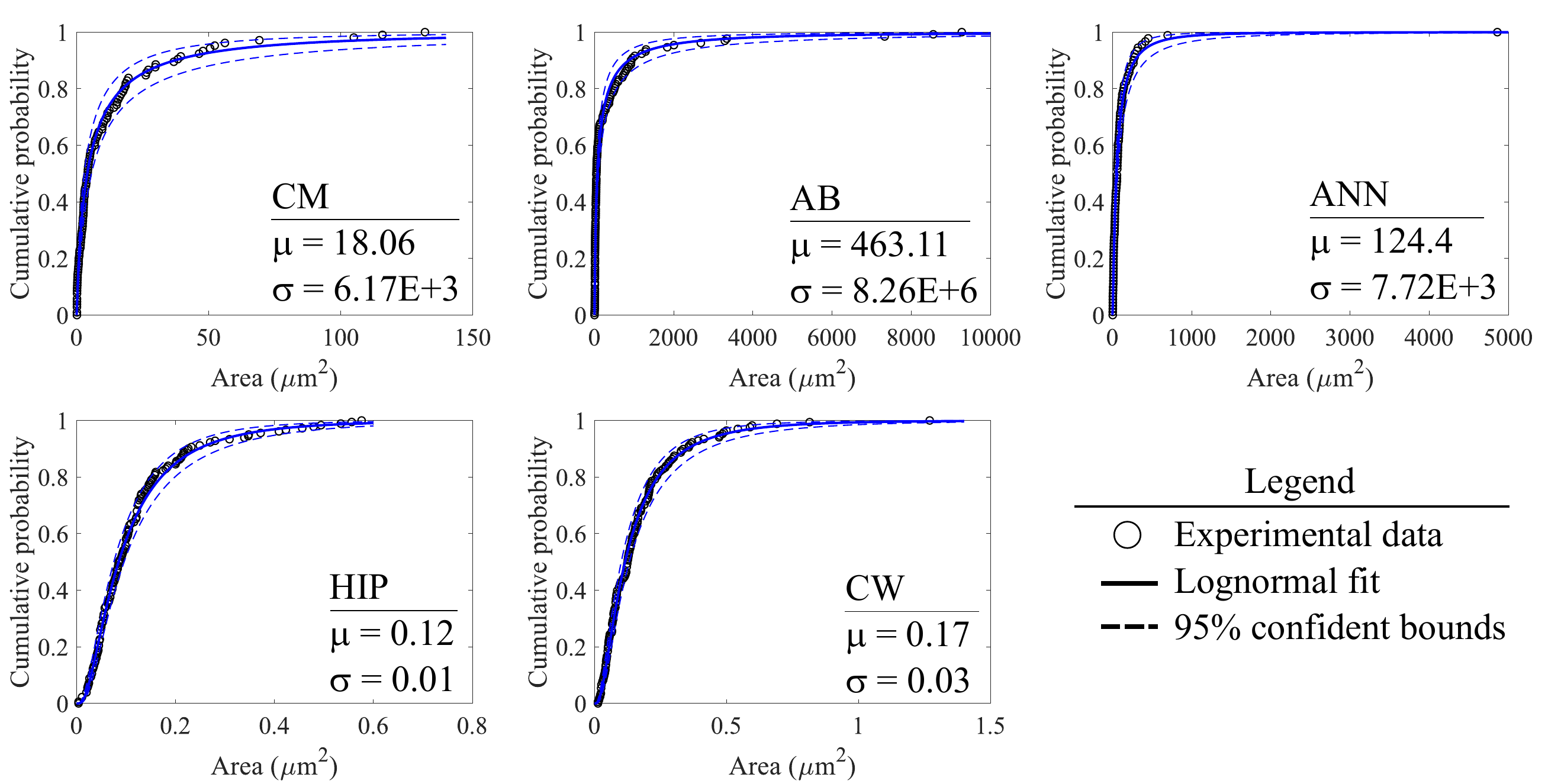}
    \caption{Cumulative distribution functions, and calculated averages and standard deviations from a lognormal distribution fit, for the pore area in each tested material condition. Note that more than 100 pores and LOF were evaluated for each case.}
    \label{fig:CDF}
\end{figure}

 To quantitatively compare the defects across the tested conditions, the cumulative distribution function (CDF) of the pore area, along with the best fit average and standard deviation assuming a lognormal distribution, for each material (more than 100 pores and LOF analyzed per condition) is shown in Figure \ref{fig:CDF}. As expected based on the micrographs, the average size and standard deviation of the pore area exhibit strong variations from condition-to-condition. For example, the porosity observed in the HIP and CW conditions exhibit areas of less than 1 $\upmu$m$^{2}$; in fact, the average area for both of these is 170 nm. Conversely, the AB and ANN conditions exhibit significantly increased pore areas, reaching 100s of $\upmu$m$^{2}$ on average, with large standard deviations consistent with the diversity of defects observed in the micrographs in Figure \ref{fig:Porosity}. Finally, it is worthwhile to mention that isolated defects were observed in the CM material. While generally small (as indicated by the average area of 18 $\upmu$m$^{2}$), some of these features were appreciable in size, resulting in a standard deviation on par with that observed for the ANN condition.

\subsection {Mechanical behavior}
\label{Sec:Materials behaviour}
\begin{figure}[hbpt]
    \centering
    \includegraphics[width=\textwidth]{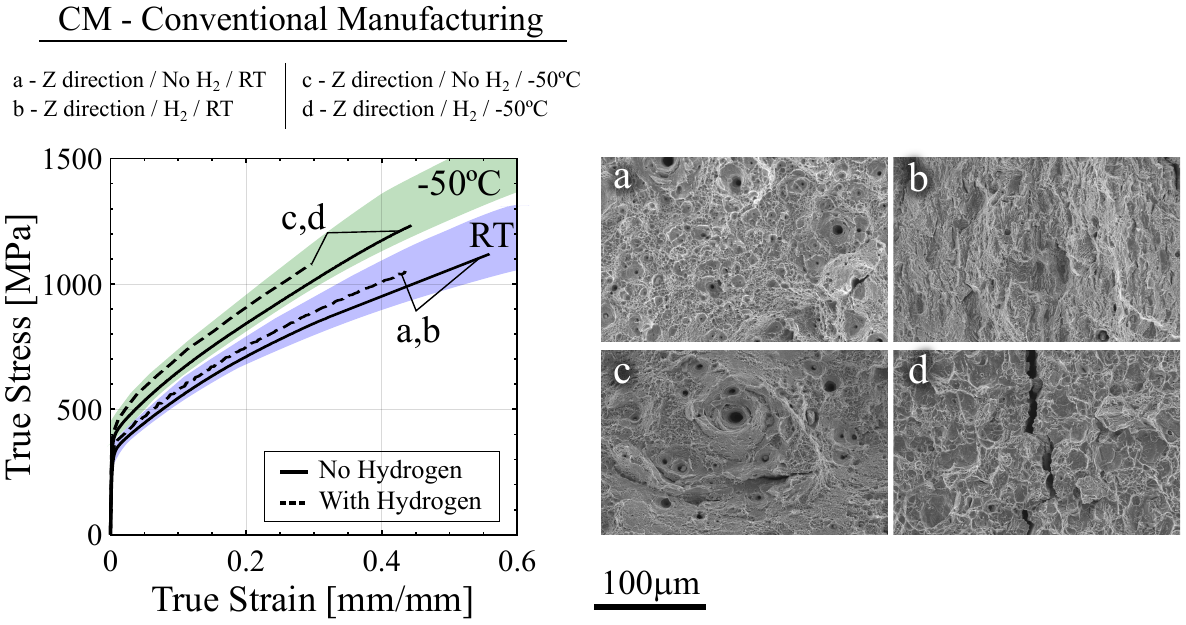}
    \caption{True stress-true strain relationships for CM 316L as a function of temperature and hydrogen content. Corresponding representative micrographs of the fracture surface for each condition are shown in (a)-(d).}
    \label{fig:NP}
\end{figure}

The true stress-true strain relationships measured for hydrogen-charged and non-charged CM 316L tested at ambient temperature (RT) and -50$^{\circ}$ C, along with representative micrographs of the fracture surface for each tested hydrogen and temperature condition, are shown in Figure \ref{fig:NP}. Note that the curves were truncated at the onset of necking as defined by Considere's criterion (where the true stress is equal to the work hardening rate \cite{Bauchau2008}) and that overview micrographs for this condition, and all other tested conditions, are provided in Appendix B. Examination of the flow curves demonstrates that at both RT and -50$^{\circ}$ C, the hydrogen-charged CM 316L exhibits an increased yield strength, but reduced uniform strain (\textit{i.e.}, strain corresponding to the onset of necking). Such results are generally consistent with the literature \cite{Abraham1994, SanMarchi2008}. Temperature was also observed to have a clear effect on the mechanical behavior, with the specimens tested at -50$^{\circ}$ C having a higher yield strength, lower uniform strain, and an increased ultimate tensile strength relative to that observed at RT. Additionally, notable differences were observed in the fractography. As shown in Figures \ref{fig:NP}a and c, the CM 316L samples strained to failure at both RT and -50$^{\circ}$ C in the absence of hydrogen exhibit widespread ductile microvoid coalescence. However, upon the introduction of hydrogen, a more brittle fracture behavior is noted. For example, hydrogen-charged CM 316L strained to failure at RT still exhibited evidence of ductile tearing, but isolated cleavage-like features were observed on the fracture surface (Figure \ref{fig:NP}b). However, these brittle cleavage features are widely observed in the hydrogen-charged CM 316L tested at -50$^{\circ}$ C, along with significant secondary cracking (Figure \ref{fig:NP}d). Such brittle features are consistent with the notable reduction in uniform strain relative to the other three temperature/hydrogen conditions, as well as with literature reports of hydrogen having a more deleterious influence on mechanical behavior in CM 316L at modest cryogenic temperatures \cite{Zhang2008,Han1998,Michler2008}.  Note that there were some differences in fracture morphology across the specimen cross-section, which are driven by the evolving complex stress state as damage nucleates, coalesces, and leads to final fracture. The regions where the above fracture morphologies were observed for all conditions in this and the following sections are highlighted in the overview images shown in Appendix B.

\begin{figure}[ht]
    \centering
    \includegraphics[width=\textwidth]{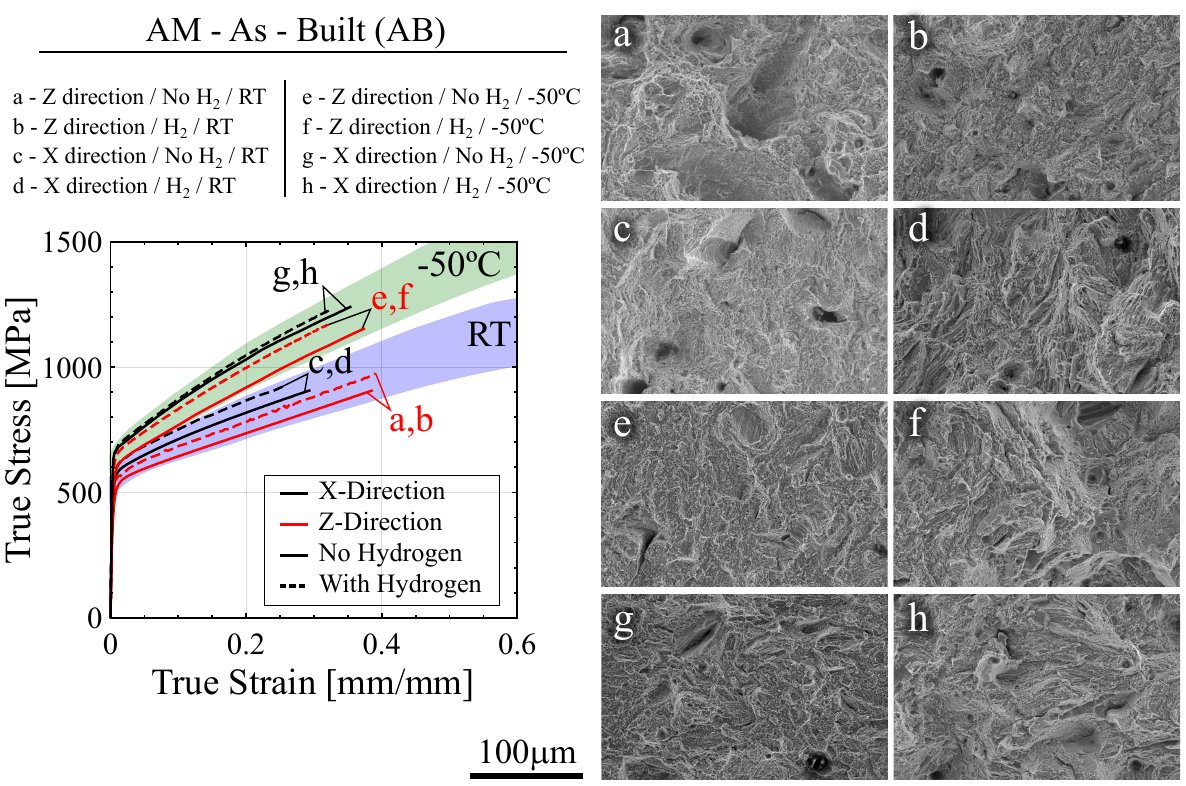}
    \caption{True stress-true strain relationships for the AB material condition as a function of directionality, temperature, and hydrogen content. Corresponding representative micrographs of the fracture surface for each condition are shown in (a)-(h).}
    \label{fig:RAW}
\end{figure}

The true stress-true strain curves for hydrogen-charged and non-charged AB 316L strained parallel and perpendicular (referred to as Z and X, respectively) to the build direction at RT and -50$^{\circ}$ C are shown in Figure \ref{fig:RAW}, along with representative fractographs for every tested combination of directionality, hydrogen charging, and temperature. As with the CM 316L, all hydrogen-charged conditions exhibit an increased yield strength relative to the non-charged condition. However, notable effects of directionality and temperature on the mechanical properties were observed for both the non-charged and hydrogen-charged AB 316L. First, consistent with the literature \cite{Dryepondt2021}, Z-oriented AB 316L exhibits increased ductility, but reduced yield strength relative to X-oriented specimens; this trend was observed for both the hydrogen-charged and the non-charged case. Regarding temperature effects, both the hydrogen-charged and non-charged specimens for each orientation exhibited increased yield strength and work hardening behavior at -50$^{\circ}$ C relative to that noted at room temperature, consistent with prior literature in CM 316L \cite{Michler2008} and AM 316L \cite{Hong2022}. Examination of the fracture surface also revealed similar fracture morphologies as reported by Hong et al. in AM 316L \cite{Hong2022}. Specifically, a largely ductile fracture surface was generally observed with isolated cleavage facets that became more frequent when hydrogen was introduced or as the temperature decreased. Intermittent features with sizes and shapes similar to the porosity noted in Figure \ref{fig:Porosity} were also noted across all tested conditions.

\begin{figure}[ht]
    \centering
    \includegraphics[width=\textwidth]{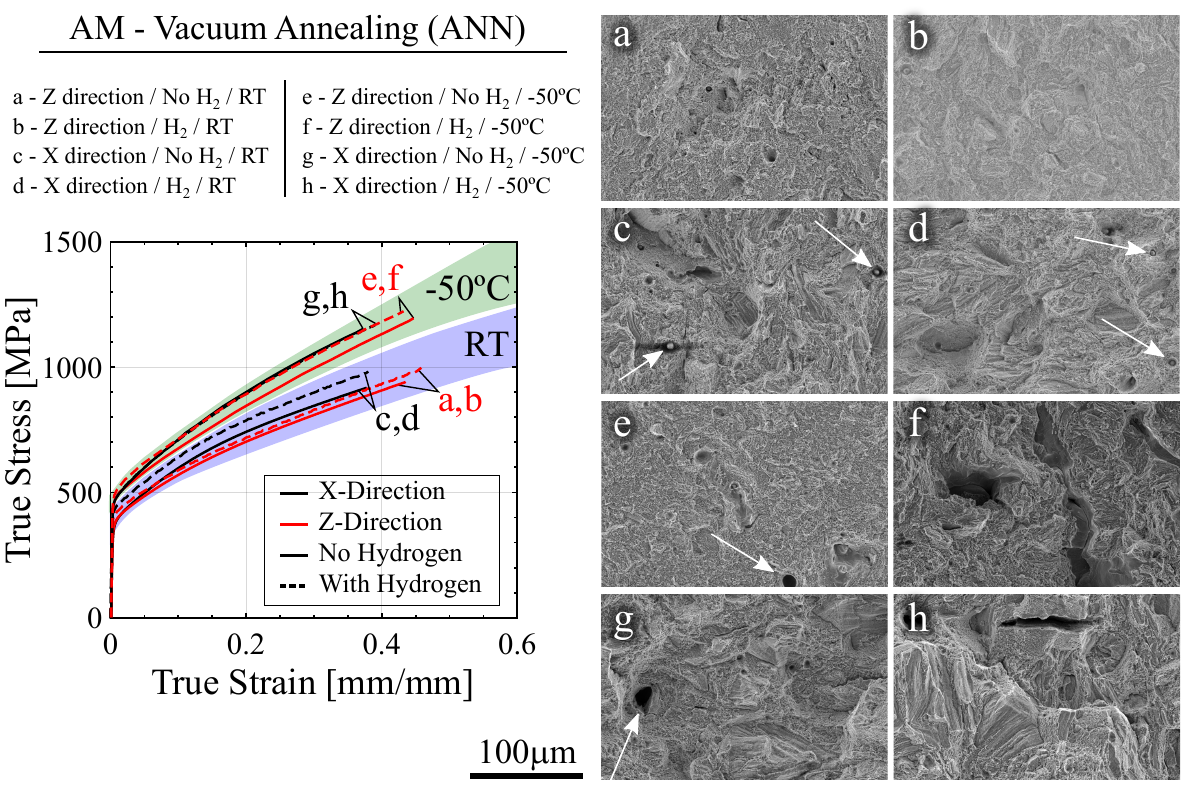}
    \caption{True stress-true strain relationships for the ANN material condition as a function of directionality, temperature, and hydrogen content. Corresponding representative micrographs of the fracture surface for each condition are shown in (a)-(h).}
    \label{fig:ANN}
\end{figure}

Figure \ref{fig:ANN} shows the true stress-true strain curves and fracture surfaces for each tested specimen for the ANN condition. As expected from literature \cite{Shamsujjoha2018}, the application of an annealing treatment reduces the yield strength but increases the ductility of AM 316L (as demonstrated by the larger uniform strain for the ANN condition). The annealing treatment also suppressed the differences in yield strength between the Z and X-oriented specimens, though the Z direction samples (red lines in Figure \ref{fig:ANN}) exhibited a higher uniform strain in all tested cases for the ANN condition. However, as was noted for the AB specimens, temperature plays an important role on the flow behavior of AM 316L, with a increased yield strength and work hardening rate observed at -50$^{\circ}$ C. While the annealing treatment clearly impacts the mechanical properties, evidence of defects from the printing process remain visible. For example, unmelted particles are shown by the white arrows in Figures \ref{fig:ANN}c-d and a large pore can be observed in Figures \ref{fig:ANN}e and g, also indicated by a white arrow. Regarding fracture morphologies, the fracture surface was again largely ductile for most tested ANN conditions, with increasing amounts of cleavage fracture features being observed as the temperature decreased and hydrogen content increased. These cleavage features are particularly notable for the hydrogen-charged, X-oriented specimen tested at -50$^{\circ}$ C (Figure \ref{fig:ANN}h), along with evidence of secondary cracking that was also observed in other ANN conditions (Figures \ref{fig:ANN}f-g).

\begin{figure}[ht]
    \centering
    \includegraphics[width=\textwidth]{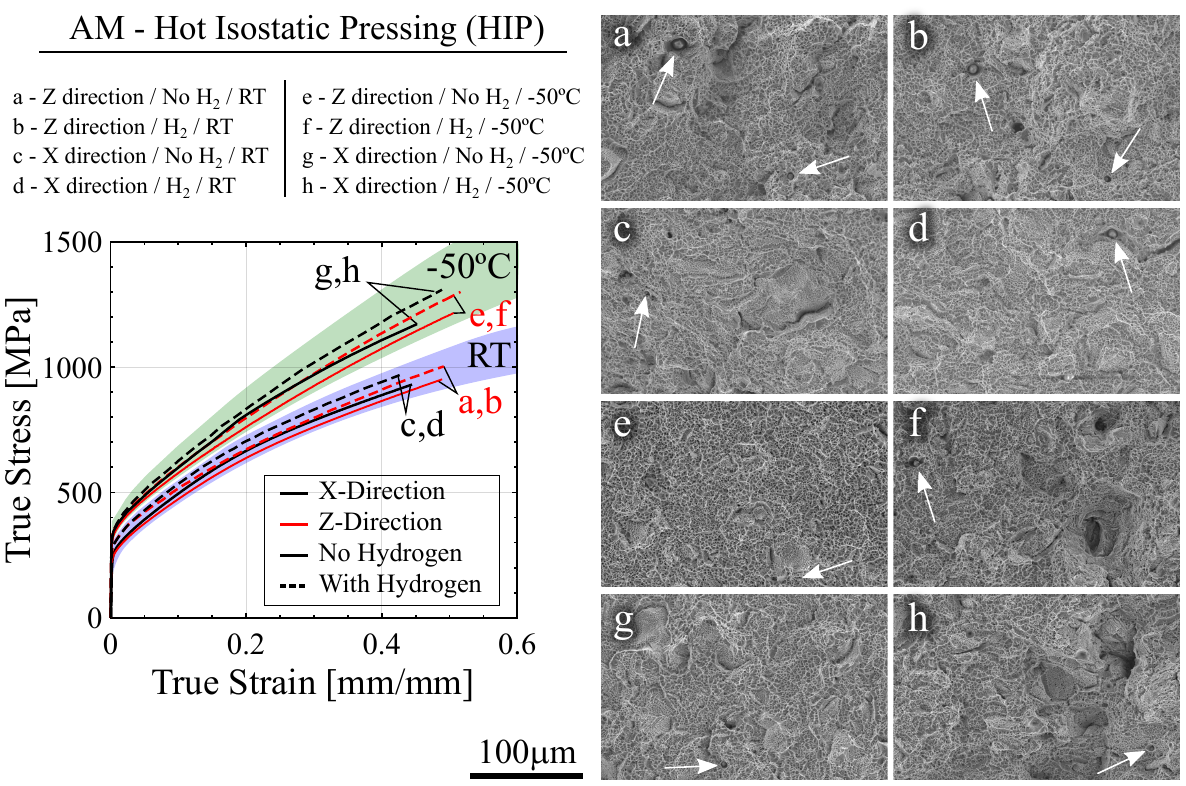}
    \caption{True stress-true strain relationships for the HIP material condition as a function of directionality, temperature, and hydrogen content. Corresponding representative micrographs of the fracture surface for each condition are shown in (a)-(h).}
    \label{fig:HIP}
\end{figure}

The true stress-true strain curves and fracture surfaces for the HIP condition are shown in Figure \ref{fig:HIP}. Two observations are immediately notable from these data: (1) the fracture morphologies are quite similar across all tested conditions, and (2) nearly all specimens are exhibiting a uniform strain of nearly 0.5 regardless of hydrogen content, direction, or temperature. Considering the former, with the HIP condition fracture surfaces are all largely composed of fracture features consistent with ductile microvoid coalescence, with intermittent evidence of cleavage facets. As with the ANN and AB conditions, these facets do seem to increase in frequency with increasing hydrogen content and decreasing temperature, but the majority of the observed area remains ductile in nature. Moreover, previously observed evidence of defects from the printing process are largely gone (some evidence of unmelted powder are noted in Figure \ref{fig:HIP}a-h; white arrows), consistent with the observed micrographs in Figure \ref{fig:Porosity}. Such results are consistent with the increased uniform strain observed across all tested HIP conditions. Regarding the flow curves, hydrogen content increases and lowering the temperature both generally increase the observed yield strength and work hardening rate. Interestingly, despite the significant processing, the Z-direction samples still exhibit slightly lower yield strength but higher ductility relative to the X-direction specimens \cite{Ulmer1991}.

\begin{figure}[htb]
    \centering
    \includegraphics[width=\textwidth]{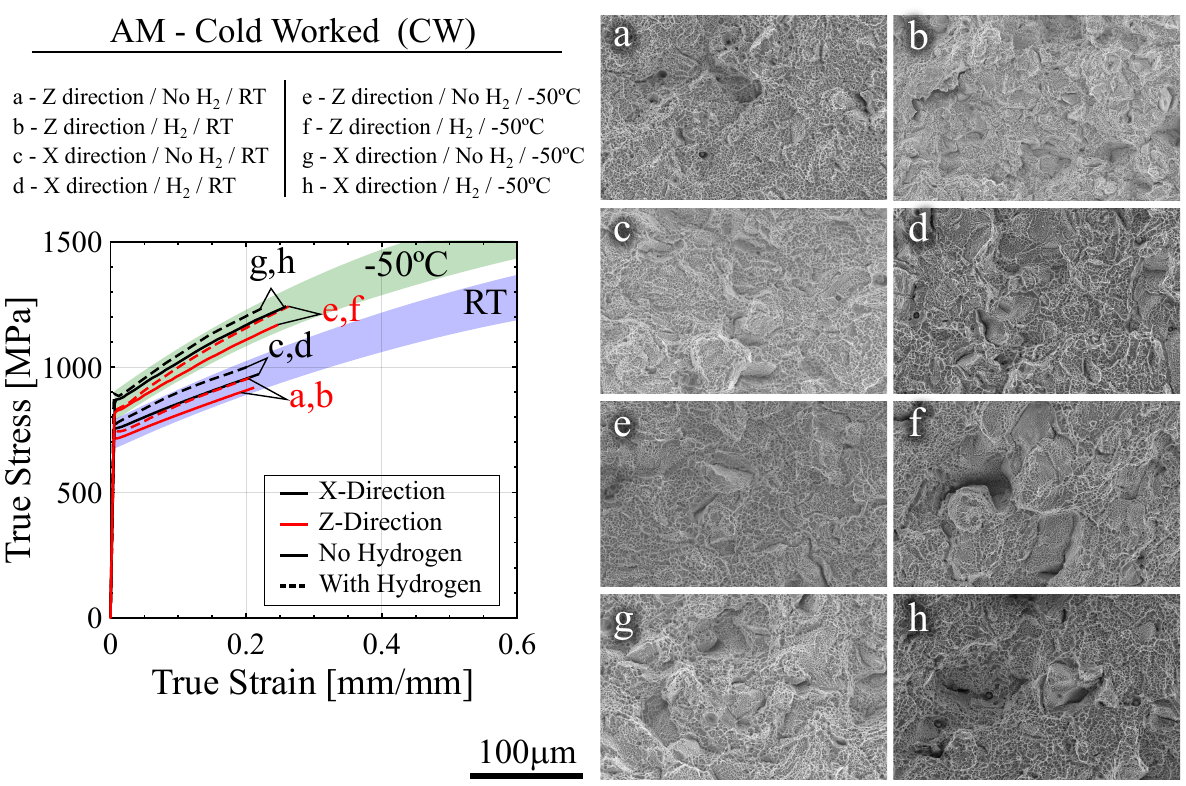}
    \caption{True stress-true strain relationships for the CW material condition as a function of directionality, temperature, and hydrogen content. Corresponding representative micrographs of the fracture surface for each condition are shown in (a)-(h).}
    \label{fig:CW}
\end{figure}

Lastly, Figure \ref{fig:CW} shows the true stress-true strain curves and fractographs for the HIP material that was subjected to 30\% pre-strain prior to hydrogen charging. This pre-strain step resulted in a notable increase in yield strength and a reduction in the uniform strain of all tested conditions. It is noteworthy that this hardening was not eliminated by the elevated temperature gaseous charging step. Consistent with the other tested material conditions, the hydrogen-charged specimens exhibit a higher yield strength and elevated work hardening rate over the analogous non-charged case. Examination of the fracture surface images reveals a generally similar morphology as was observed for the HIP condition in Figure \ref{fig:HIP}. Specifically, evidence of ductile microvoiding is widely observed across all tested conditions along with intermittent cleavage facets. Qualitatively, it appears that there is an increase in the number of cleavage facets for a given testing condition relative to the analogous HIP specimen, regardless of temperature, hydrogen content, or direction. Such results are consistent with previous reports of an increase in brittle fracture morphologies for pre-strained 316L, especially after hydrogen charging \cite{Zhou2019,Wang2019} or deformation at low temperatures \cite{Michler2015}.

\section{Discussion}
\label{Sec:Discussion}

The preceding results demonstrate that the mechanical behavior and fracture surface morphology of AM 316L is dependent on testing temperature, sample orientation, and hydrogen content. Critically, these data also reveal that the deformation behavior of AM 316L after hydrogen charging can be strongly modified by post-processing. In the following discussion, the results of this study are evaluated in the context of the literature. The basis for the observed similar or even improved hydrogen compatibility of post-processed AM 316L relative to CM 316L is then assessed, with specific focus on the role of strain-induced martensite. The discussion then concludes with a brief overview of the engineering implications of this study.

\subsection {Comparison of results with literature }
\label{Sec:Comparison Results}

\begin{figure}[pt]
    \centering
    \includegraphics[height=0.90\textheight]{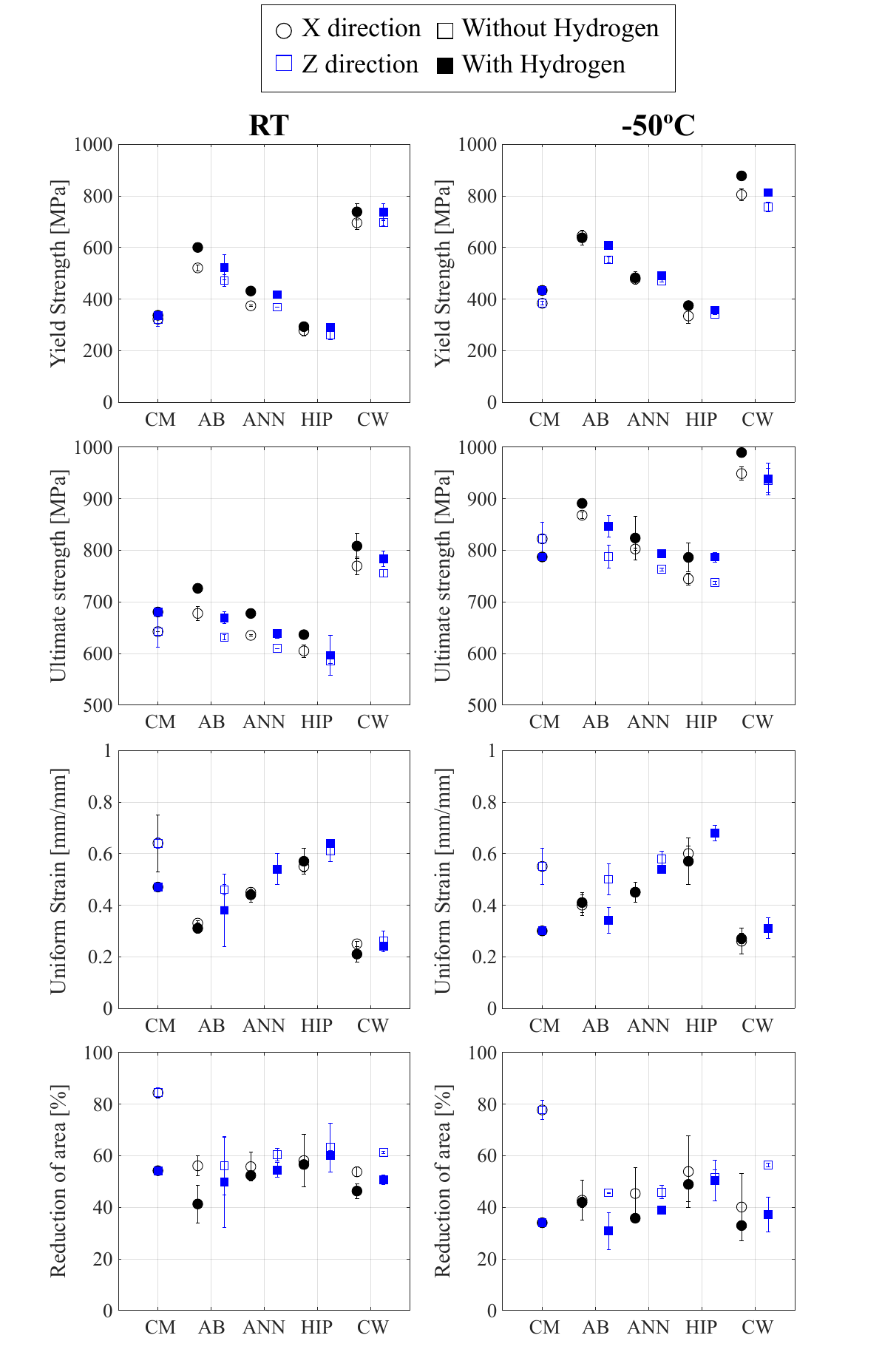}
    \caption{Measured yield strength, ultimate tensile strength, uniform strain, and reduction of area for all tested AM and CM conditions as a function of temperature and hydrogen content. Error bars correspond to the standard deviation of the measured metric from duplicate experiments.}
    \label{fig:Props}
\end{figure}

A summary of the mechanical properties obtained for the preceding experiments is presented in Figure \ref{fig:Props}, with specific focus on the effect of direction, temperature, and post-processing treatment on the yield strength, ultimate tensile strength, uniform strain, and reduction of area. Multiple aspects of the current study are consistent with expectations established by prior literature. For example, while the specific magnitude may vary from study to study, an increase in yield strength with hydrogen content has been broadly reported in the literature for 316L and is generally linked to hydrogen providing a solid solution strengthening contribution \cite{Abraham1994, Ulmer1991, SanMarchi2008}. A similar effect of hydrogen increasing the ultimate tensile strength has also been reported \cite{SANMARCHI2021, Komatsu2021}. While these prior studies have largely been focused on CM 316L, the same general trends have been reported in the literature for AM 316L. For example, Park et al. reported an increase in the hardness of LPBF 316L after hydrogen charging \cite{Park2021}, Bertsch et al. reported higher yield strength and ultimate tensile strength for both DED and LPBF 316L \cite{Bertsch2021}, and Claeys et al. noted an increase in yield strength for electrochemically-charged AM 316L \cite{CLAEYS2023}. Interestingly, a recent study by Hong et al. reported similar yield strength for hydrogen-charged and non-charged LPBF 316L, but given the sample dimensions and employed charging temperature and time, the test coupons may not have had a spatially uniform hydrogen concentration, which may have affected this result \cite{Hong2022}. Alternatively, this also could be due to the reduced hydrogen concentration present given the use of a lower hydrogen gas pressure during charging. 

\begin{figure}[!htb]
    \centering
    \includegraphics[width=\textwidth]{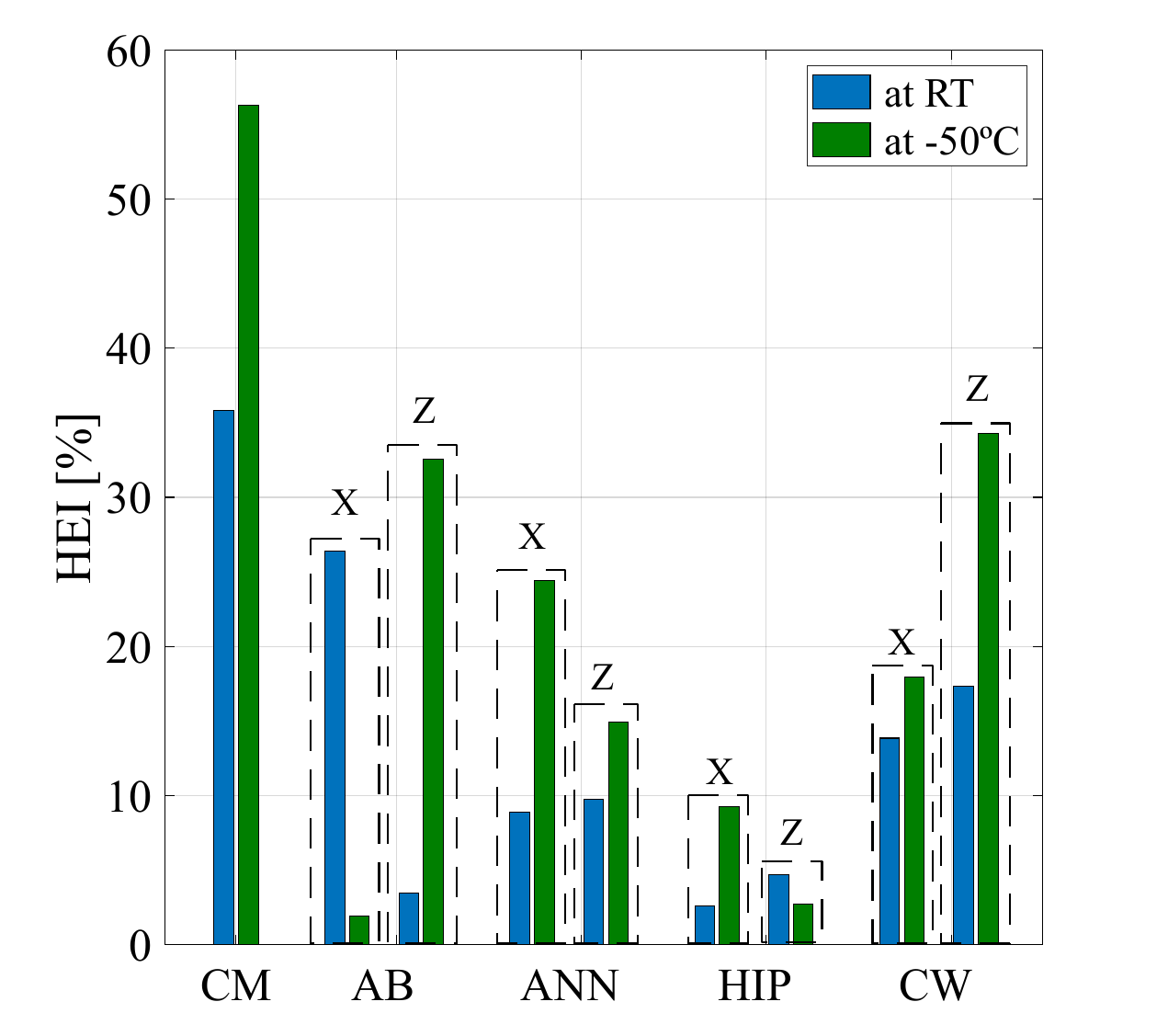}
    \caption{Calculated HEI values for each AM and CM 316L condition at both room temperature and -50$^{\circ}$ C.}
    \label{fig:HEI}
\end{figure}

 Regarding ductility measurements, Figure \ref{fig:Props} demonstrates that the CM 316L exhibits significantly higher ductility (as quantified by reduction in area) relative to all tested AM conditions in the absence of hydrogen at both room temperature and -50$^{\circ}$ C. Given that porosity is observed in all AM conditions, even after HIP treatment (Figure \ref{fig:Porosity}), it is postulated that these defects cause an inherent reduction in the ductility of the AM materials relative to the CM 316L condition \cite{Grech2022}. However, upon hydrogen charging, all AM 316L and CM 316L conditions exhibit a decrease in the reduction of area, though the relative decrease is most clearly observed in the CM 316L condition, particularly at -50$^{\circ}$ C. While the AB condition remains less ductile than the CM 316L after hydrogen charging at both test temperatures, post-processing appears to have a significant influence on relative performance. Specifically, the ANN condition exhibits a reduction of area on par with CM 316L at room temperature and a slightly higher ductility at -50$^{\circ}$ C; this improvement in the ductility with annealing of AB 316L was also noted by Bertsch et al. for DED and LPBF 316L \cite{Bertsch2021,Yin2019,Urabinczyk2022}. However, the most interesting result is the performance of the HIP 316L material, which has a modestly increased reduction of area relative to CM 316L at room temperature, but nearly double the reduction of area of CM 316L at -50$^{\circ}$ C. This result suggests that the HIP condition is more hydrogen-resistant, despite the porosity that acts to decrease the hydrogen-free ductility of the material. The increased resistance of the HIP material can be further demonstrated through a comparison of the calculated HEI (Equation \ref{eqn:HEI}) for all tested material conditions and test temperatures, shown in Figure \ref{fig:HEI}. While HEI-based evaluations can be misleading (HEI can be higher, but the hydrogen-affected reduction of area may still be high), these results clearly show that hydrogen has a minimal effect on the HIP condition. In fact, while there is certainly an effect of temperature and directionality, these results suggest that all AM conditions are affected less by hydrogen than CM 316L relative to their respective hydrogen-free condition. Given the nominally similar hydrogen concentrations present in all tested materials, this implies that an intrinsic feature of the AM materials is lowering the influence of hydrogen. 

\subsection {Role of strain-induced martensite}
\label{Sec:Martensite}
The preceding section suggests that all AM conditions are less affected by hydrogen than CM 316L, especially at -50$^{\circ}$ C. Moreover, by applying common post-processing treatments that attenuate the influence of AM-induced defects or the complex AB grain structure on the ductility, this intrisincally improved resistance can be harnessed to create materials that have higher hydrogen compatibility than CM 316L. However, in order to better optimize AM post-processing treatments, it is necessary to understand the basis for the improved hydrogen resistance. Numerous literature reports have noted that (1) strain-induced martensite (SIM) can significantly degrade the hydrogen resistance of CM austenitic stainless steels \cite{SANMARCHI2021, Han1998, Zhang2008} and (2) there is a generally reduced propensity for SIM formation in AM austenitic stainless steels \cite{Baek2017, Hong2022}. The increased degradation of the hydrogen-charged CM 316L's performance at -50$^{\circ}$ C relative to that observed at room temperature (and the reduced extent of such effects in the AM conditions) implies a role of SIM, but quantification is necessary to support this postulation. One way to assess this possibility is to calculate the Md$_{30}$ temperature \cite{Hong2022,Metalnikov2022,Angel1954}, which is a composition-based metric that can be used to compare the stability of the austenite phase against the transformation to $\alpha$'-martensite:
\begin{equation}
\label{eqn:MD30}
\text{Md}_{30}=413-462(\text{wt}\%[\text{C}+\text{N}])-9.2(\text{wt}\%\text{Si})-8.1(\text{wt}\%\text{Mn})-13.7(\text{wt}\%\text{Cr})-9.5(\text{wt}\%\text{Ni})-18.5(\text{wt}\%\text{Mo})
\end{equation}

\begin{figure}[!htb]
    \centering
    \includegraphics[width=\textwidth]{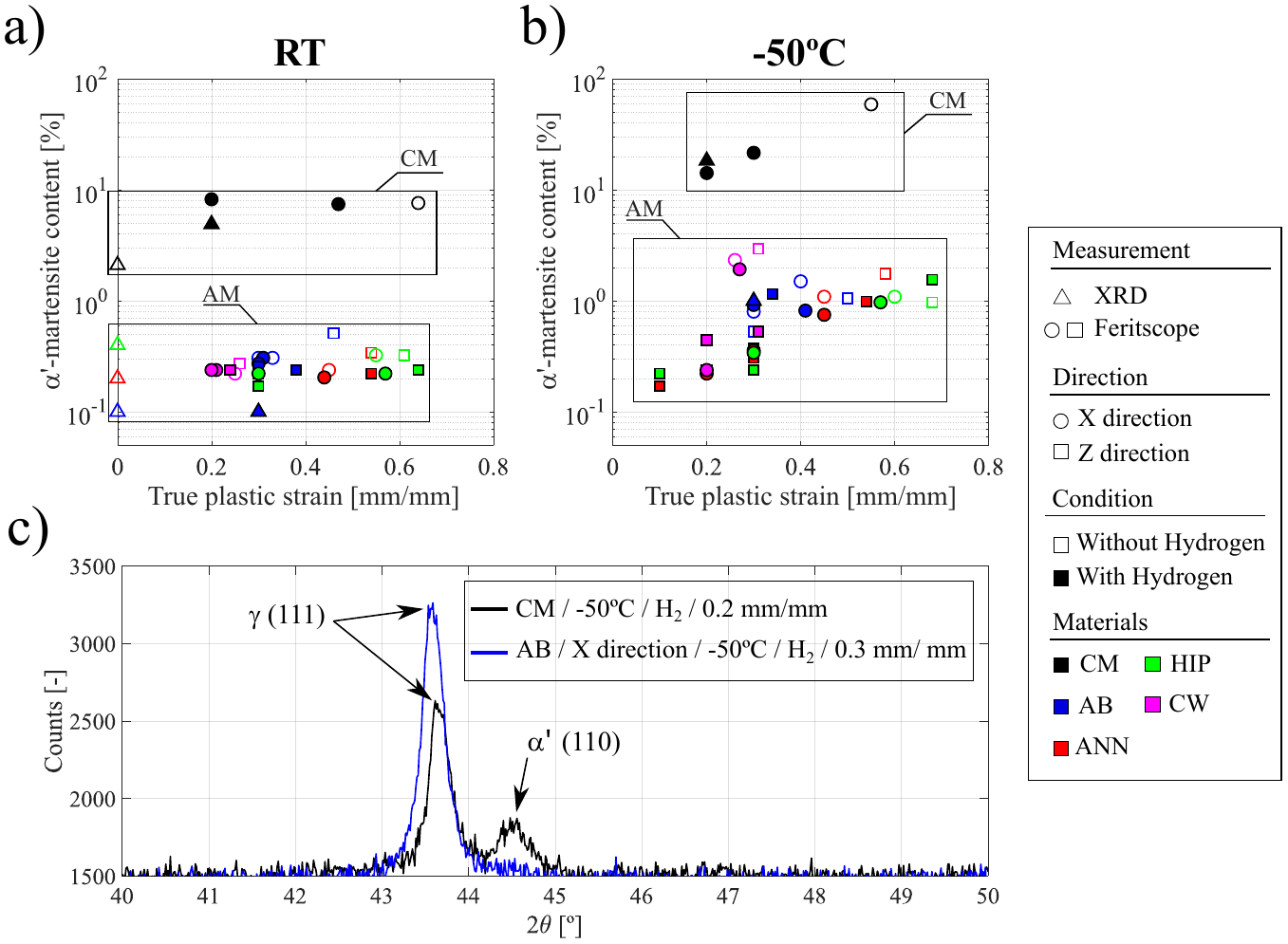}
    \caption{Measured $\alpha$'-martensite volume fraction as a function of true plastic strain for all tested AM and CM 316L conditions at a) RT and b) -50$^{\circ}$ C (note error bars are not shown as they were smaller than the size of the plotted symbols); c) XRD diffraction patterns indicating profile differences between high and low $\alpha$'-martensite content (for XRD measurements the analysed data were always in the X direction)}
    \label{fig:Feritscope}
\end{figure}

Using the compositions listed in the Methods section, this equation results in an Md$_{30}$ temperature of $\approx$14.2 and $\approx$33.4$^{\circ}$ C for the AM and CM 316L materials. This result strongly suggests that the AM material should be inherently more resistant to the formation of $\alpha$'-martensite during deformation, which aligns well with the observed hydrogen susceptibility data in Figure \ref{fig:Props}. However, Hong et al. \cite{Hong2022} demonstrated that SIM formation depends on both the phase stability and the applied mechanical driving force (\textit{i.e.}, plastic strain). Therefore, to better understand how SIM evolves in the different conditions, magnetic-based measurements of the $\alpha$'-martensite volume fraction were made as a function of plastic strain using a Feritscope for specimens deformed at both room temperature and -50$^{\circ}$ C. These data were augmented by XRD measurements and subsequent Rietveld refinement-based calculations of the martensite phase fraction. The results of these analyses, shown in Figure \ref{fig:Feritscope}, strongly imply a linkage between the reduced propensity for SIM formation and a decreased effect of hydrogen on the AM material ductility. Specifically, the $\alpha$'-martensite fraction for hydrogen-charged AM materials tested at room temperature remains less than 0.3\% for all four tested AM conditions up to true plastic strains of $>$0.6. Conversely, under the same testing conditions and true plastic strain range, the CM 316L exhibits a more than 30-fold higher $\alpha$'-martensite fraction. This difference becomes even more substantial when deformation takes place at -50$^{\circ}$ C, with the CM 316L reaching an $\alpha$'-martensite fraction of $>$30\% at a true plastic strain where the AM materials still exhibit less than 1\% $\alpha$'-martensite. Such differences are clearly observed when the XRD data are directly compared, as in Figure \ref{fig:Feritscope}c, where the CM condition exhibits a well-defined $\alpha$'-martensite peak that is not present in the AB condition tested to an even higher plastic strain level. The parameter Ni$_{\text{eq}}$ can also be used to assess austenite stability (see \cite{Yamabe2017}). However, differences in the value of Ni$_{\text{eq}}$ for the AM and CM materials are small (within 1\%).

This apparent influence of limited SIM formation is consistent with the findings of other authors studying hydrogen effects in AM austenitic stainless steels. For example, Baek et al. reported the absence of SIM in hydrogen-charged AM 304L even after 50\% strain \cite{Baek2017}. An absence of SIM was also reported in LPBF 316L by Khaleghifar et al. after electrochemical charging and tensile deformation reaching $>$60\% strain \cite{Khaleghifar2021} as well as by Kong et al. after  electrochemical charging alone \cite{Kong2020}. From the microstructural perspective, Hong et al. noted a preference for deformation twinning over SIM formation in hydrogen-charged LPBF 316L at room temperature \cite{Hong2022}, consistent with prior results for hydrogen-free AM 316L \cite{Shamsujjoha2018}. Similar ideas were also proposed by Metalkinov et al., who suggested that the better resistance of the AB 316L to hydrogen embrittlement relative to CM 316L was due to the unique dislocation cell structures of AM material \cite{Wang2019} and lower $\alpha$'-martensite content \cite{Metalnikov2022}. Moreover, Hong et al. also reported the onset of SIM at -40$^{\circ}$ C in hydrogen-charged LPBF 316L \cite{Hong2022}, but it was highly localized to twin-twin or twin-grain boundary interactions, consistent with the low reported SIM fraction values after cryogenic testing in Figure \ref{fig:Feritscope}. Interestingly, these authors suggested that annealing treatments could be used to further suppress the onset of SIM in AM 316L, which is consistent with the observed  lower $\alpha$'-martensite fraction for the ANN condition relative to the AB condition for a given true plastic strain in Figure \ref{fig:Feritscope}. However, what remains unclear is the microstructural basis for the observed further improvement of the HIP material over the ANN condition in the presence of hydrogen. As shown in Figure \ref{fig:EBSD}, the HIP material does  exhibit an equiaxed grain structure, while the ANN material has a grain structure closer to that observed in the AB condition. It is possible that the more uniform grain structure suppresses localization of deformation, further hindering the onset of SIM and improving the hydrogen resistance. This would be consistent with the need for higher plastic strains in the HIP condition to obtain the same $\alpha$'-martensite fraction present in the ANN material, but targeted near-crack characterization efforts are necessary to confirm this hypothesis \cite{Harris2020,Martin2020}.

\subsection {Implications of results}
\label{Sec:Implications}
Given the specialized nature of many components employed in hydrogen transportation, storage, and distribution infrastructure, there is growing interest in utilizing AM to support the proliferation of the hydrogen economy. However, as has been documented widely in the literature \cite{Gordon2020}, AM materials are prone to intrinsic defects that degrade their fracture resistance, which has historically hindered their use in fracture-critical or higher-risk structural components. Moreover, regarding hydrogen applications, these defects (such as pores and LOF) seem to contribute to the increased hydrogen-assisted crack growth rates measured in AM alloys relative to incumbent wrought materials \cite{Shoemaker2022}. While cost considerations motivate the use of AM materials in the `as-built' condition, additional thermal processing steps could be justified if they are shown to yield AM alloy performance on par with incumbent materials for the application of interest. As the present study sought to directly assess the influence of post-processing on hydrogen compatibility of AM 316L, it is therefore useful to comment on the implications of the present results in the broader context of adopting AM for manufacturing components for hydrogen service.

First, the present results demonstrate that LPBF-fabricated AM 316L stainless steel subjected to common post-build thermal treatments can exhibit similar or even improved resistance to hydrogen embrittlement relative to CM 316L (Figure \ref{fig:Props}). For example, both the ANN and HIP conditions had RA values similar to or slightly higher than those measured for CM316L after hydrogen charging and testing at room temperature. The hydrogen performance of the post-processed AM 316L was even further improved relative to CM 316L at -50$^{\circ}$ C, with the RA of the HIP condition reaching nearly twice that of the CM 316L. Interestingly, the CW AM condition also exhibits similar RA as the CM 316L at both room temperature and -50$^{\circ}$ C, despite having a significantly higher yield strength. These data suggest that AM 316L could possibly be used for components in hydrogen transportation, storage, and distribution infrastructure that are typically fabricated from CM 316L. There is still significant work needed to enable this possibility; for example, it is important to evaluate the effect of surface conditions (as the current study was conducted on machined specimens) that are likely to be present on AM-built components on AM 316L hydrogen compatibility. However, these initial results are promising as the ability to leverage AM and common post-processing methods to fabricate hydrogen-compatible parts on-demand would be beneficial given the inherently distributed nature of hydrogen fuel infrastructure.

Second, regarding the post-processing steps employed, it is notable that the improved hydrogen performance in the AM 316L was observed without any optimization of the post-processing parameters. For example, the grain structure of the ANN still exhibited significant remnants of the AB microstructure; it is possible that the annealing parameters could be optimized to promote a more equiaxed grain structure that may have better hydrogen compatibility. While the HIP-treated material clearly performed the best of the AM 316L conditions, this post-processing strategy is the most expensive to execute and requires special equipment, which will hinder broader implementation. Therefore, if optimization of the more economical annealing treatment can result in similar performance as the current HIP condition, this would likely accelerate adoption of AM for hydrogen service. 

Finally, the current study evaluated the LPBF approach for AM, but there are a number of other AM techniques that should be assessed from a hydrogen compatibility perspective. For example, metal binder jet printing can print 316L components but with notable benefits relative to LPBF: (1) lower residual stresses due to the absence of rapid solidification found in LPBF, (2) materials are printed with isotropic properties since the component is not directionally solidified, and (3) the printing process itself is faster and does not require inert atmosphere \cite{Mostafaei2021}. However, to date, there is little to no work studying hydrogen effects in materials fabricated using AM methods outside of DED and LPBF \cite{Khaleghifar2021,CLAEYS2023, Maksimkin2022, Bertsch2021}. As each given AM approach offers specific advantages and disadvantages, it would be worthwhile to evaluate the hydrogen compatibility of `best-case' specimens fabricated using these other methods.

\section{Conclusions}
\label{Sec:Concluding remarks}

The effect of post-processing treatment on the hydrogen embrittlement susceptibility of AM 316L stainless steel was assessed via tensile testing of hydrogen precharged and non-charged specimens at both ambient temperature and -50$^{\circ}$ C. Results were compared to companion testing executed on CM 316L and the relative performance of each condition was correlated with Feritscope-based measurements of the $\alpha'$-martensite volume fraction. From these data, the following conclusions can be made:

\begin{itemize}
    \item At ambient temperature, hydrogen charging increased the yield strength and reduced the ductility of as-built AM 316L. Experiments on AB condition also showed a sensitivity to the build direction that persisted after hydrogen charging, with a higher yield strength reported in the X direction and a higher ductility observed for the Z direction.
    
    \item Non-charged CM 316L tested at -50$^{\circ}$ C exhibited an increased ductility as compared to AB AM 316L. However, after hydrogen charging, the AB AM 316L had similar or even superior ductility relative to CM 316L.
    
    \item As reported in literature \cite{Shamsujjoha2018}, the application of post-processing strategies such as annealing or HIP treatments improved the ductility of the non-charged AM 316L relative to the AB condition at both ambient temperature and -50$^{\circ}$ C. This same improvement in ductility was also noted after hydrogen charging for the ANN and HIP AM 316L conditions. The ductility of the AM 316L conditions (regardless of charging condition) were not significantly affected by the reduction in temperature, while the CM 316L was.
    
    \item Hydrogen-charged ANN and HIP AM 316L exhibited similar ductility as the CM 316L at ambient temperature and significantly improved ductility at -50$^{\circ}$ C. The hydrogen-charged HIP AM 316L condition exhibited an RA that was nearly twice that of hydrogen-charged CM 316L at -50$^{\circ}$ C. The 30\% cold-worked AM 316L was found to have similar ductility at -50$^{\circ}$ C to CM 316L after hydrogen charging, despite having a 2-fold higher yield strength.
    
    \item The improved performance of post-processed AM 316L relative to CM 316L is tentatively attribute to the reduced propensity for strain-induced martensite formation, which has been reported to reduce hydrogen compatibility \cite{Kong2020,Metalnikov2022}. Feritscope-based measurements performed as a function of plastic strain revealed that AM 316L exhibited significantly lower strain-induced martensite at both low temperatures and after significant post-processing than CM 316L. 
    
    \item These results suggest that, with targeted post-processing, AM 316L could be leveraged for components used in hydrogen transportation, storage, and distribution applications with similar expected performance as CM 316L. Moreover, if applications require exposure to cryogenic temperatures, the present results suggest that post-processed AM 316L may exhibit improved hydrogen compatibility over CM 316L.
 
\end{itemize}

\section*{Acknowledgements}

The authors acknowledged helpful discussions and support (AM materials) from the US Army Research Laboratory (Robert Drummond, Brandon McWilliams). In addition, in regards to the HIP treatment, the authors acknowledge funding from the Henry Royce Institute PhD Access Scheme and support from Vahid Nekouie (University of Sheffield).
E. Mart\'{\i}nez-Pa\~neda additionally acknowledges financial support from the EPSRC (grants EP/V04902X/1 and EP/R010161/1) and from UKRI's Future Leaders Fellowship programme [grant MR/V024124/1]. G. \'{A}lvarez acknowledges the Principality of Asturias for the support received with the Severo Ochoa grant (PA-20-PF-BP19-087), BritishSpanish Society (Plastic Energy Award) and Margarita Salas Postdoctoral contract (Ref.: MU-21-UP2021-030) funded by the University of Oviedo through the Next Generation European Union.

\appendix
\section*{Appendix A. Verification of the influence of precharging conditions}
\label{Sec:Appendix A}

To confirm that the thermal charging procedure would not tangibly modify the properties of the AB material, an AB specimen was exposed to the same conditions (270$^{\circ}$ C for 400 hours) as the hydrogen-charged samples, but in a laboratory air environment. As shown in the figure below, the flow curves of the non-charged AB material bound the measured curve for the AB sample that was subjected to the additional heat treatment. Therefore, these data strongly suggest that the thermal treatment used for hydrogen charging the specimens did not induce appreciable changes in the mechanical behavior of the AB material.

\bigskip

\begin{figure}[H]
    \centering
    \includegraphics[width=9cm]{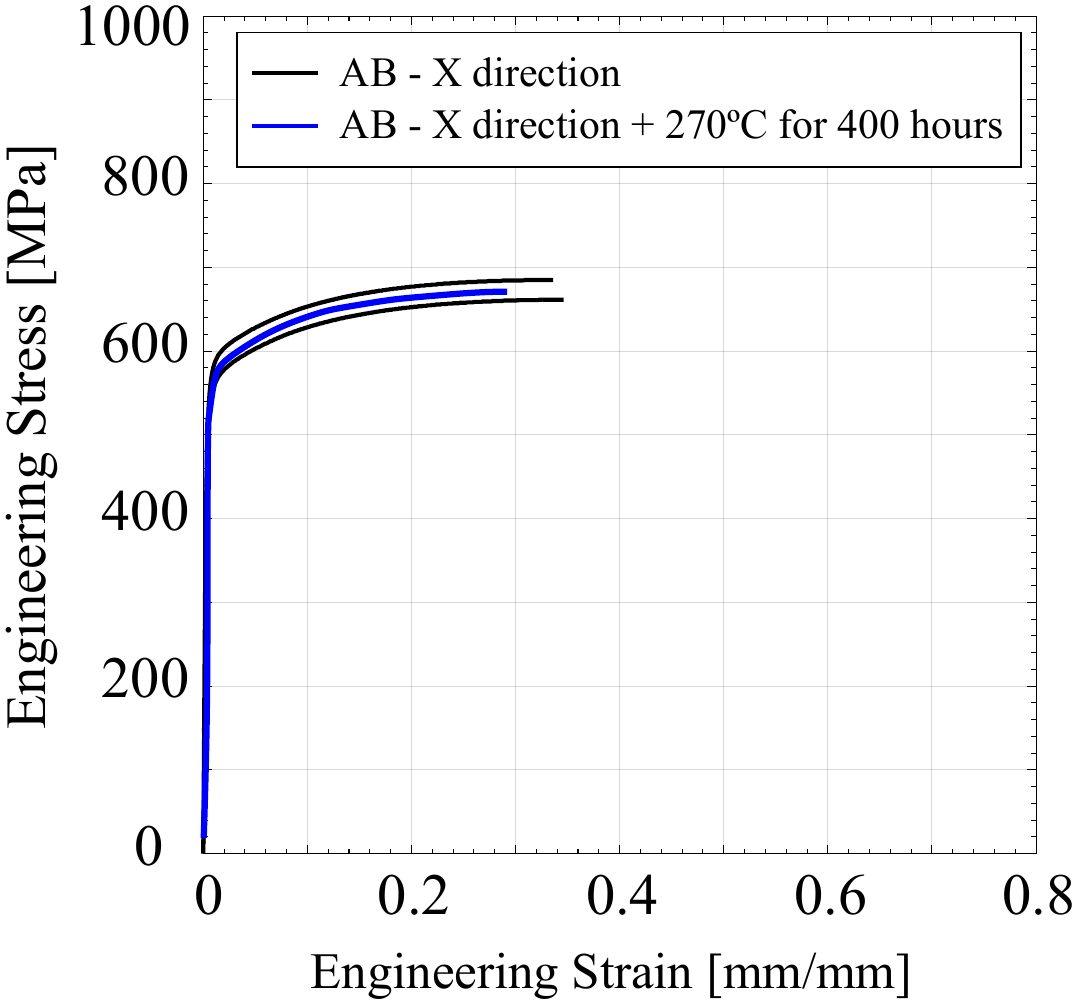}
    \caption{Stress versus strain curves showing how the conditions of the hydrogen charging step (400 h and 270$^\circ$ C) do not lead to significant changes in mechanical behaviour. The results compare a test on AB samples and AB samples that have been heat treated in a furnace for 400 h at 270$^\circ$ C.}
    \label{fig:Appendix A}
\end{figure}

\section*{Appendix B. Fractographic analysis of the samples for all conditions}
\label{Sec:Appendix B}

To enhance the analysis of the fracture surfaces with a greater number of magnifications, this appendix collects several images for all testing conditions and materials. For each case, a low magnification image is provided, with a blue shadow representing the area where the fracture morphology is uniform. Additionally, a close-up detail of this zone is included alongside each image to complement the fractography described in the text.

\begin{figure}[H]
    \centering
    \includegraphics[width=\textwidth]{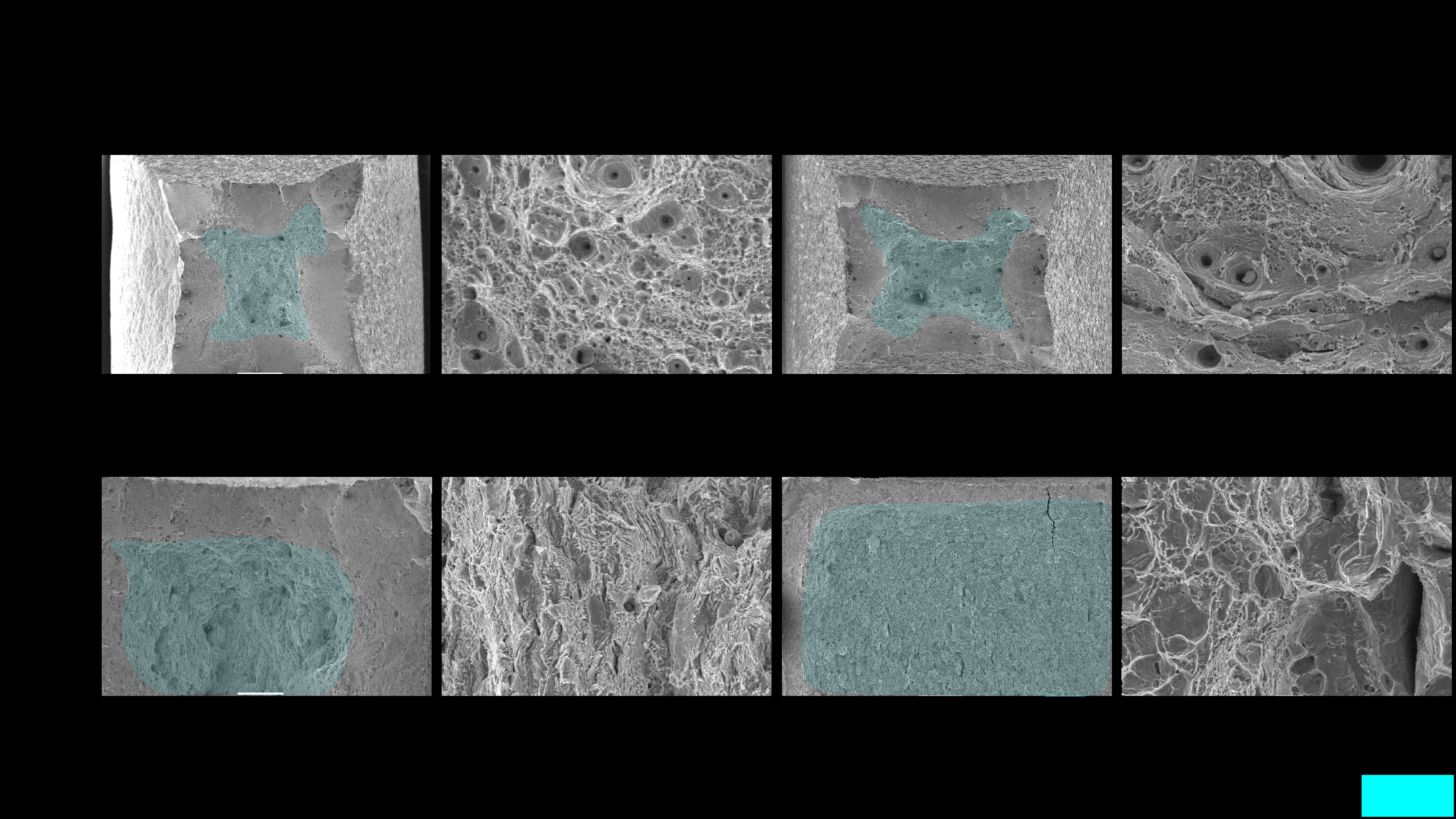}
    \caption{Fractographs of the tensile surfaces of the CM material after being tested.}
    \label{fig:Appendix A}
\end{figure}

\begin{figure}[H]
    \centering
    \includegraphics[width=\textwidth]{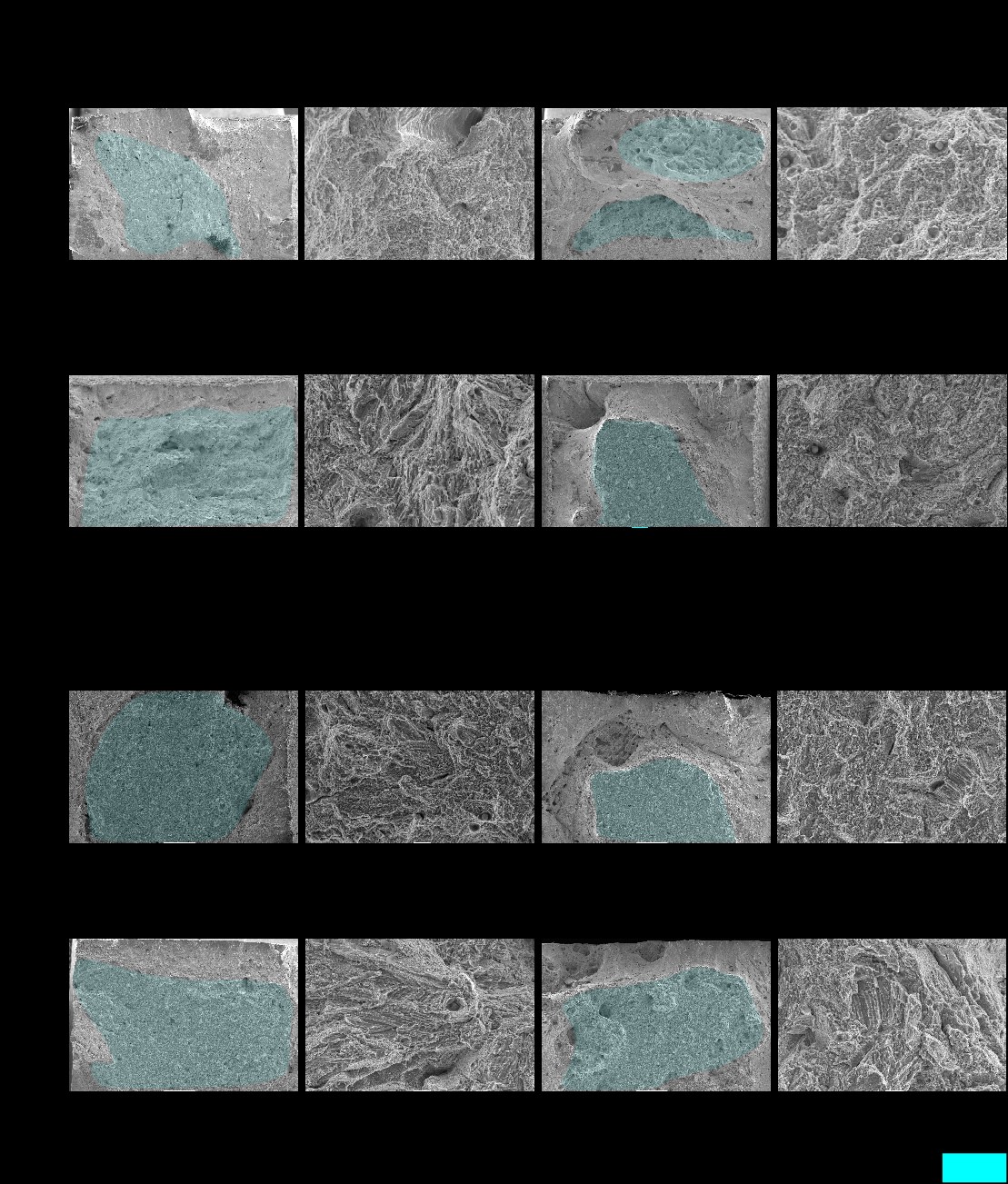}
    \caption{Fractographs of the tensile surfaces of the AB material after being tested.}
    \label{fig:Appendix A}
\end{figure}

\begin{figure}[H]
    \centering
    \includegraphics[width=\textwidth]{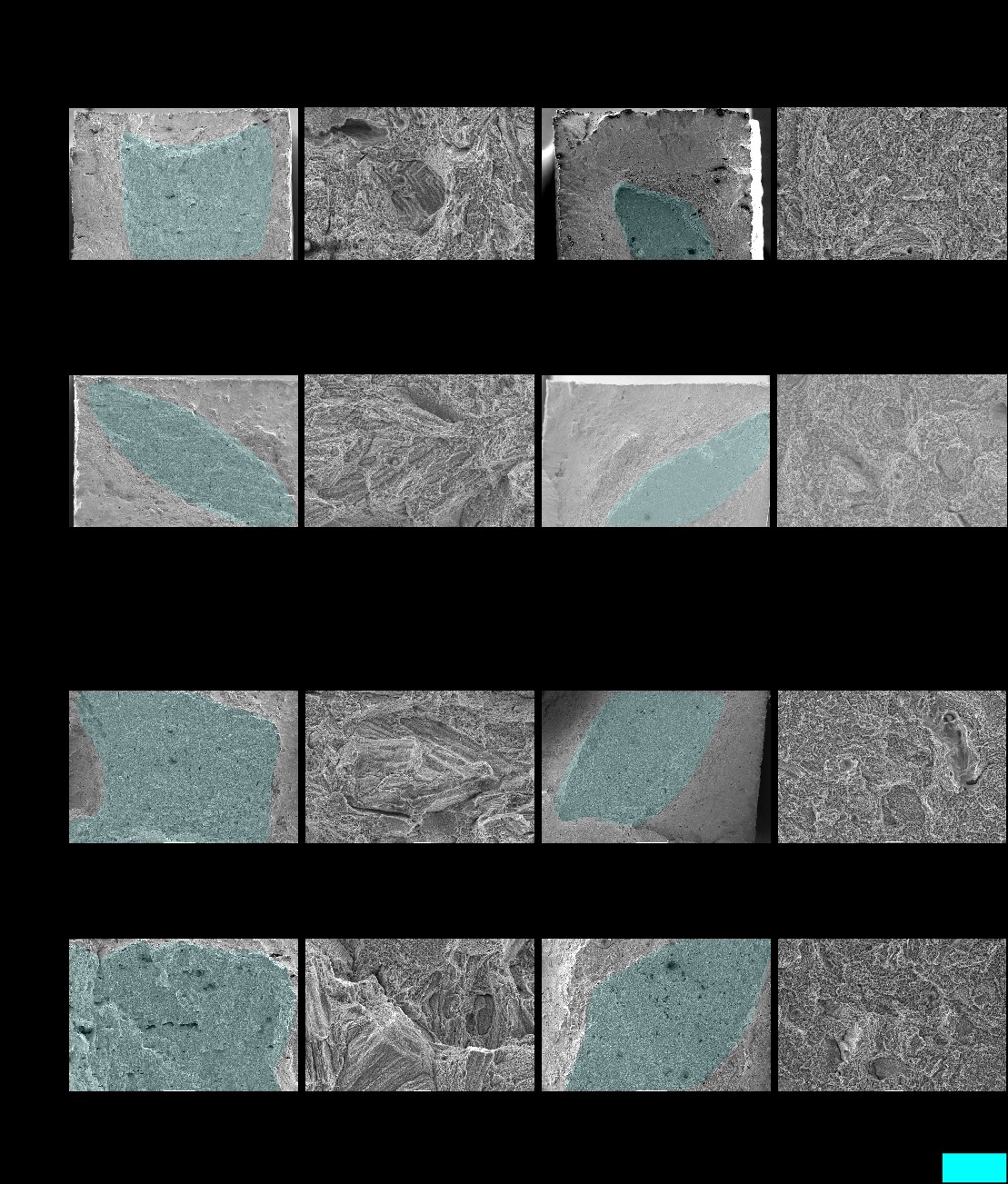}
    \caption{Fractographs of the tensile surfaces of the ANN material after being tested.}
    \label{fig:Appendix A}
\end{figure}

\begin{figure}[H]
    \centering
    \includegraphics[width=\textwidth]{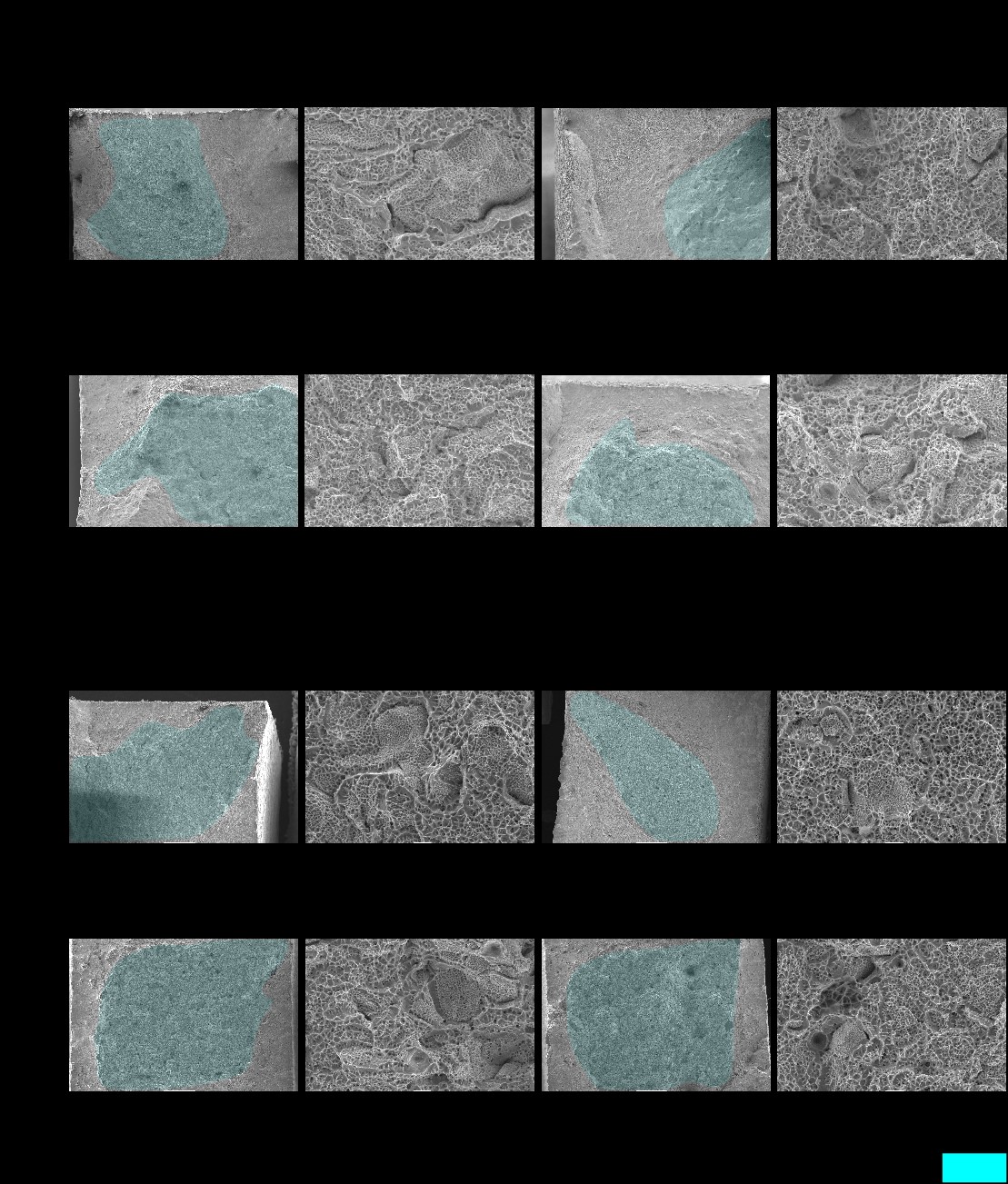}
    \caption{Fractographs of the tensile surfaces of the HIP material after being tested.}
    \label{fig:Appendix A}
\end{figure}

\begin{figure}[H]
    \centering
    \includegraphics[width=\textwidth]{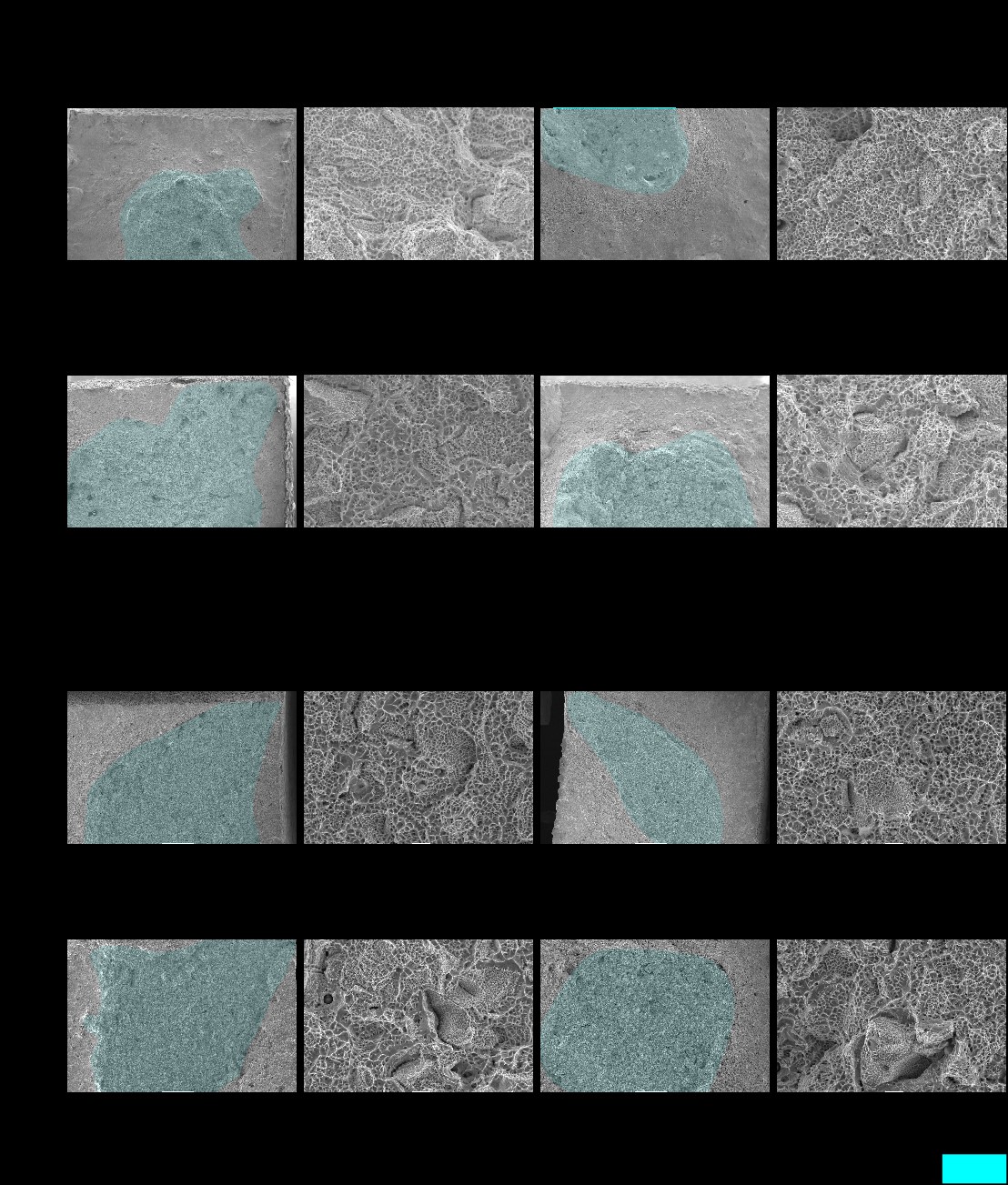}
    \caption{Fractographs of the tensile surfaces of the CW material after being tested.}
    \label{fig:Appendix A}
\end{figure}

\section*{Appendix C. Assessing porosity in different planes}
\label{Sec:Appendix C}

To improve the porosity characterization and highlight the differences between rolling or printing planes, this appendix collects SEM images of the porosity for both X-Y and X-RD/BD planes for each material. Additional statistics for the employed build parameters, powder source, and type of machine (Renishaw AM400) can be found in other publications, such as reported work from Argonne National Laboratory \cite{Chen2023}

\begin{figure}[H]
    \centering
    \includegraphics[width=\textwidth]{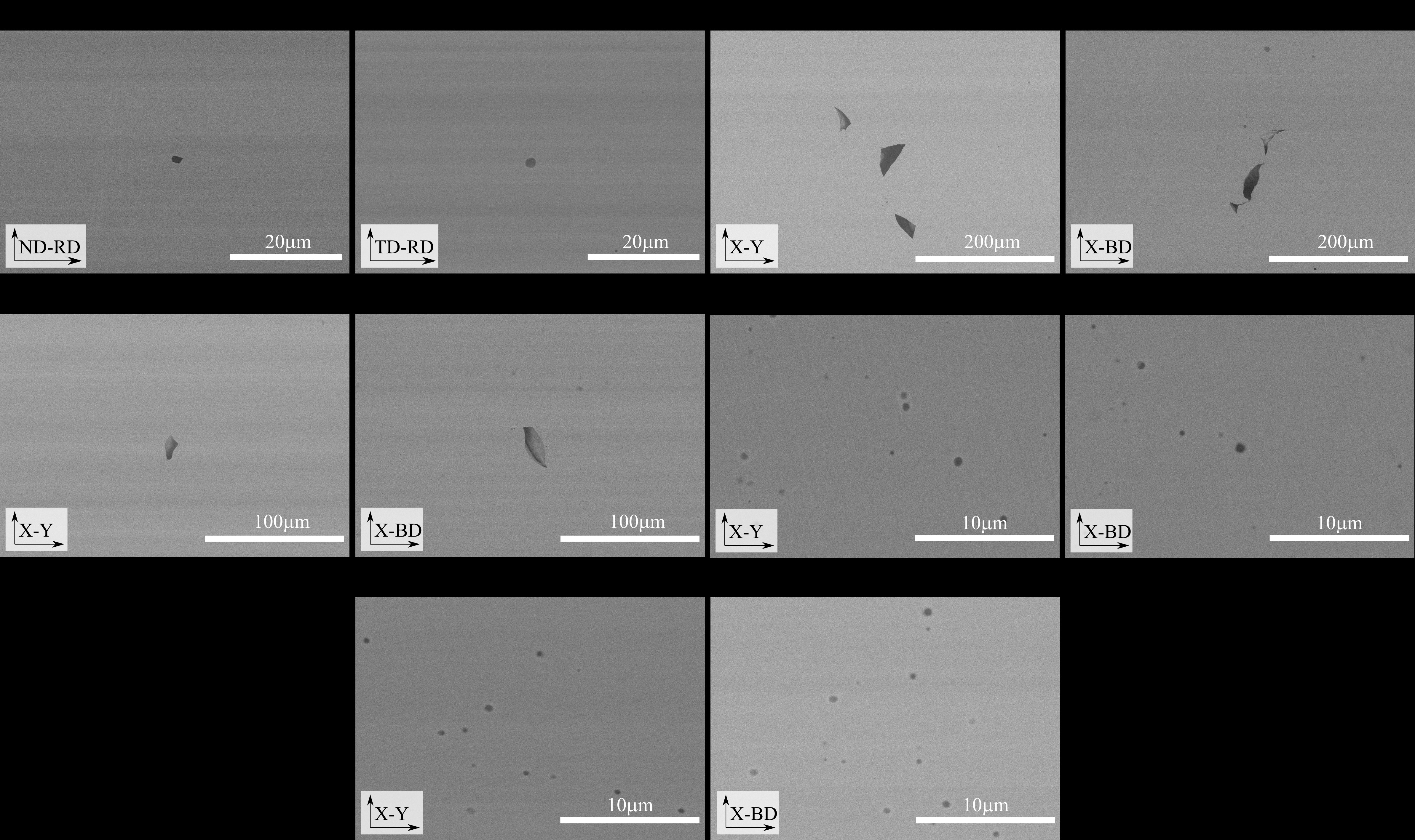}
    \caption{Comparison of porosity for all materials for X-Y and X-RD/BD planes.}
    \label{fig:Appendix C}
\end{figure}



\end{document}